\documentclass[5p,times]{elsarticle}
\usepackage{hyperref}

\journal{Knowledge Based Systems}

\usepackage{balance}
\usepackage{listings}
\usepackage{color}

\definecolor{dkgreen}{rgb}{0,0.6,0}
\definecolor{gray}{rgb}{0.5,0.5,0.5}
\definecolor{mauve}{rgb}{0.58,0,0.82}

\makeatletter
\def\lst@makecaption{%
  \def\@captype{table}%
  \@makecaption
}
\makeatother

\usepackage{tikz}

\usepackage{amssymb}
\usepackage{pifont}

\usepackage{algorithmic}        
\usepackage[vlined,ruled]{algorithm2e}    
\SetKwProg{Fn}{Function}{}{}

\graphicspath{{img/}}

\usepackage{url}
\usepackage{multirow}

\usepackage{amssymb}
\usepackage{amsmath}
\usepackage[caption=false,font=footnotesize,labelfont=sf,textfont=sf]{subfig}
\usepackage{dblfloatfix}
\usepackage[export]{adjustbox}

\usepackage{algorithmic}
\usepackage{array}

\usepackage[caption=false,font=footnotesize,labelfont=sf,textfont=sf]{subfig}

\begin{document}

\begin{frontmatter}

\title{dyngraph2vec: Capturing Network Dynamics using Dynamic Graph Representation Learning}

\author{Palash~Goyal}
\address{University of Southern California, Information Sciences Institute\\4676 Admiralty Way, Suite 1001. Marina del Rey, CA. 90292, USA}
\author{Sujit~Rokka~Chhetri}
\address{University of California-Irvine\\Irvine, CA. 92697, USA}
\author{Arquimedes Canedo}
\address{Siemens Corporate Technology\\755 College Rd E, Princeton, NJ. 08540, USA}

\begin{abstract}
Learning graph representations is a fundamental task aimed at capturing various properties of graphs in vector space. The most recent methods learn such representations for static networks. However, real-world networks evolve over time and have varying dynamics. Capturing such evolution is key to predicting the properties of unseen networks. To understand how the network dynamics affect the prediction performance, we propose an embedding approach which learns the structure of evolution in dynamic graphs and can predict unseen links with higher precision. Our model, \textit{dyngraph2vec}, learns the temporal transitions in the network using a deep architecture composed of dense and recurrent layers. We motivate the need for capturing dynamics for the prediction on a toy data set created using stochastic block models. We then demonstrate the efficacy of dyngraph2vec over existing state-of-the-art methods on two real-world data sets. We observe that learning dynamics can improve the quality of embedding and yield better performance in link prediction.
\end{abstract}


\end{frontmatter}

\section{Introduction}\label{sec:introduction}
Understanding and analyzing graphs is an essential topic that has been widely studied over the past decades. Many real-world problems can be formulated as link predictions in graphs~\cite{Gehrke2003,freeman2000visualizing,theocharidis2009network,goyal2018recommending}.
For example, link prediction in an author collaboration network~\cite{Gehrke2003} can be used to predict potential future author collaboration. Similarly, new connections between proteins can be discovered using protein interaction networks~\cite{pavlopoulos2008survey}, and new friendships can be predicted using social networks~\cite{wasserman1994social}. Recent work on obtaining such predictions use graph representation learning. These methods represent each node in the network with a fixed dimensional embedding and map link prediction in the network space to the nearest neighbor search in the embedding space~\cite{goyal2017graph}. It has been shown that such techniques can outperform traditional link prediction methods on graphs~\cite{Grover2016,Ou2016}.

\begin{figure}
\centering
  \includegraphics[width=0.35\textwidth]{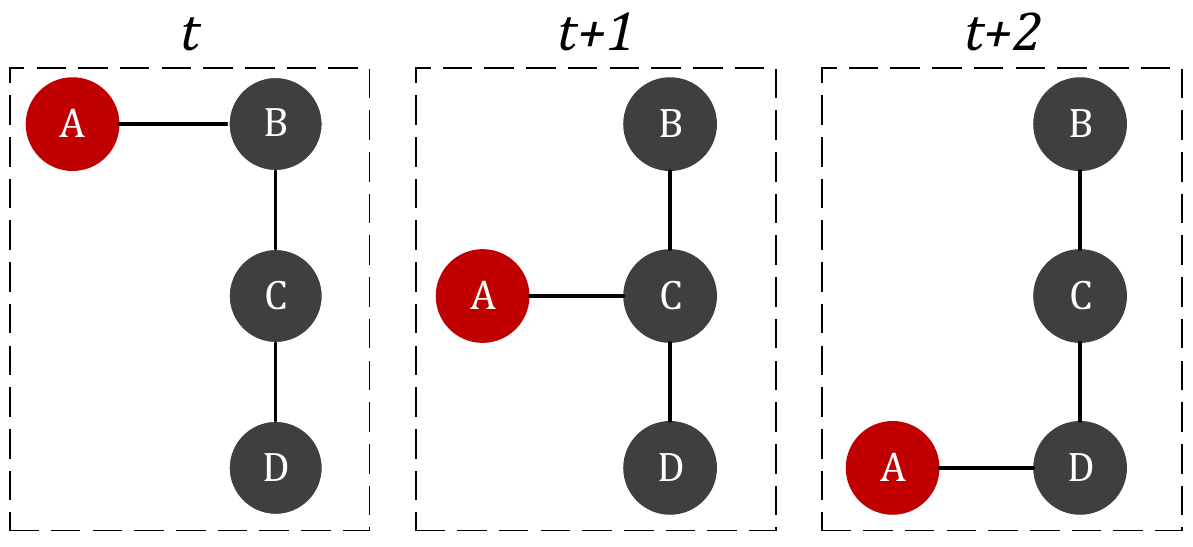}
  \vspace{-1em}
  \caption{User A breaks ties with his friend at each time step and befriends a friend of a friend. Such temporal patterns require knowledge across multiple time steps for accurate prediction.}
  \label{fig:intro_example}
  \vspace{-.6cm}
\end{figure}
Existing works on graph representation learning primarily focus on static graphs of two types: (i) aggregated, consisting of all edges until time $T$; and (ii) snapshot, which comprise of edges at the current time step $t$. These models learn latent representations of the static graph and use them to predict missing links~\cite{Ahmed2013,Perozzi2014,Cao2015,Tang2015,Grover2016,Ou2016,goyal2018embedding}.
However, real networks often have complex dynamics which govern their evolution.
As an illustration, consider the social network shown in Figure~\ref{fig:intro_example}. In this example, user A moves from one friend to another in such a way that only a friend of a friend is followed and making sure not to befriend an old friend.
Methods based on static networks can only observe the network at time $t+1$ and cannot ascertain if A will befriend B or D in the next time step.
Instead, observing multiple snapshots can capture the network dynamics and predict A's connection to D with high certainty.

In this work, we aim to capture the underlying network dynamics of evolution.
Given temporal snapshots of graphs, our goal is to learn a representation of nodes at each time step while capturing the dynamics such that we can predict their future connections.
Learning such representations is a challenging task.
Firstly, the temporal patterns may exist over varying period lengths.
For example, in Figure~\ref{fig:intro_example}, user A may hold to each friend for a varying $k$ length.
Secondly, different vertices may have different patterns.
In Figure~\ref{fig:intro_example}, user A may break ties with friends whereas other users continue with their ties.
Capturing such variations is extremely challenging.
Existing research builds upon simplified assumptions to overcome these challenges.
Methods including DynamicTriad~\cite{zhou2018dynamic}, DynGEM~\cite{goyal2018dyngem} and TIMERS~\cite{zhang2017timers} assume that the patterns are of short duration (length 2) and only consider the previous time step graph to predict new links.
Furthermore, DynGEM and TIMERS make the assumption that the changes are smooth and use a regularization to disallow rapid changes.

In this work, we present a model which overcomes the above challenges.
\textit{dyngraph2vec} uses multiple non-linear layers to learn structural patterns in each network.
Furthermore, it uses recurrent layers to learn the temporal transitions in the network.
The look back parameter in the recurrent layers controls the length of temporal patterns learned.
We focus our experiments on the task of link prediction.
We compare dyngraph2vec with the state-of-the-art algorithms for dynamic graph embedding and show its performance on several real-world networks including collaboration networks and social networks. 
Our experiments show that using a deep model with recurrent layers can capture temporal dynamics of the networks and significantly outperform the state-of-the-art methods on link prediction. {\color{black}We emphasize that our work is targeted towards link prediction and not node classification.} {\color{black} Furthermore, our algorithm works on both aggregated and snapshot temporal graphs.}

Overall, our paper makes the following contributions:
\begin{enumerate}
    \item We propose \textit{dyngraph2vec}, a dynamic graph embedding model which captures temporal dynamics.
    \item We demonstrate that capturing network dynamics can significantly improve the performance on link prediction.
    \item We present variations of our model to show the key advantages and differences.
    \item We publish a library, DynamicGEM~\footnote{https://github.com/palash1992/DynamicGEM}, implementing the variations of our model and state-of-the-art dynamic embedding approaches.
\end{enumerate}


\section{Related Work}\label{sec:related}
Graph representation learning techniques can be broadly divided into two categories: (i) static graph embedding, which represents each node in the graph with a single vector; and (ii) dynamic graph embedding, which considers multiple snapshots of a graph and obtains a time series of vectors for each node.
Most analysis has been done on static graph embedding.
Recently, however, some works have been devoted to studying dynamic graph embedding.

\subsection{Static Graph Embedding}
Methods to represent nodes of a graph typically aim to preserve certain properties of the original graph in the embedding space.
Based on this observation, methods can be divided into (i) distance preserving, and (ii) structure preserving.
Distance preserving methods devise objective functions such that the distance between nodes in the original graph and the embedding space have similar rankings.
For example, Laplacian Eigenmaps~\cite{belkin2001laplacian} minimizes the sum of the distance between the embeddings of neighboring nodes under the constraints of translational invariance, thus keeping the nodes close in the embedding space.
Similarly, Graph Factorization~\cite{Ahmed2013} approximates the edge weight with the dot product of the nodes' embeddings, thus preserving distance in the inner product space.
Recent methods have gone further to preserve higher order distances.
Higher Order Proximity Embedding (HOPE)~\cite{Ou2016} uses multiple higher-order functions to compute a similarity matrix from a graph's adjacency matrix and uses Singular Value Decomposition (SVD) to learn the representation.
GraRep~\cite{Cao2015} considers the node transition matrix and its higher powers to construct a similarity matrix.

On the other hand, structure-preserving methods aim to preserve the roles of individual nodes in the graph.
\emph{node2vec}~\cite{Grover2016} uses a combination of breadth-first search and depth-first search to find nodes similar to a node in terms of distance and role.
Recently, deep learning methods to learn network representations have been proposed. These methods inherently preserve the higher order graph properties including distance and structure.
SDNE~\cite{Wang2016}, DNGR~\cite{cao2016deep} and VGAE~\cite{kipf2016variational} use deep autoencoders for this purpose.
Some other recent approaches use graph convolutional networks to learn inherent graph structure~\cite{kipf2016semi,bruna2013spectral,henaff2015deep}.

\begin{figure*}[!hb]
    \centering
    \begin{adjustbox}{minipage=\linewidth,scale=0.95}
    \vspace{-1.5em}
    \subfloat[DynGEM]{\label{fig_m2} \includegraphics[width=0.32\textwidth]{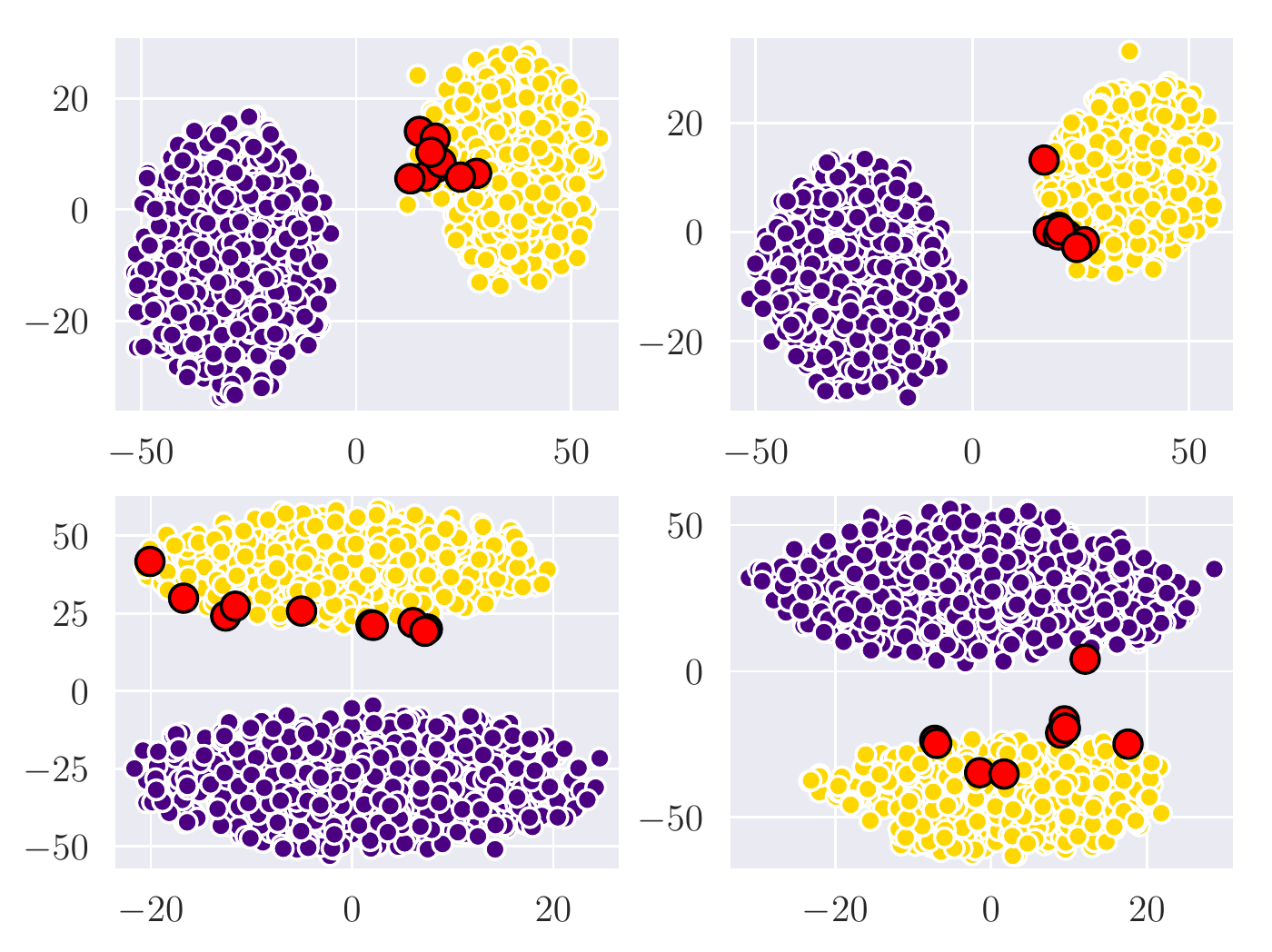}}
    \hfil
    \subfloat[\textit{optimalSVD}]{\label{fig_m4} \includegraphics[width=0.32\textwidth]{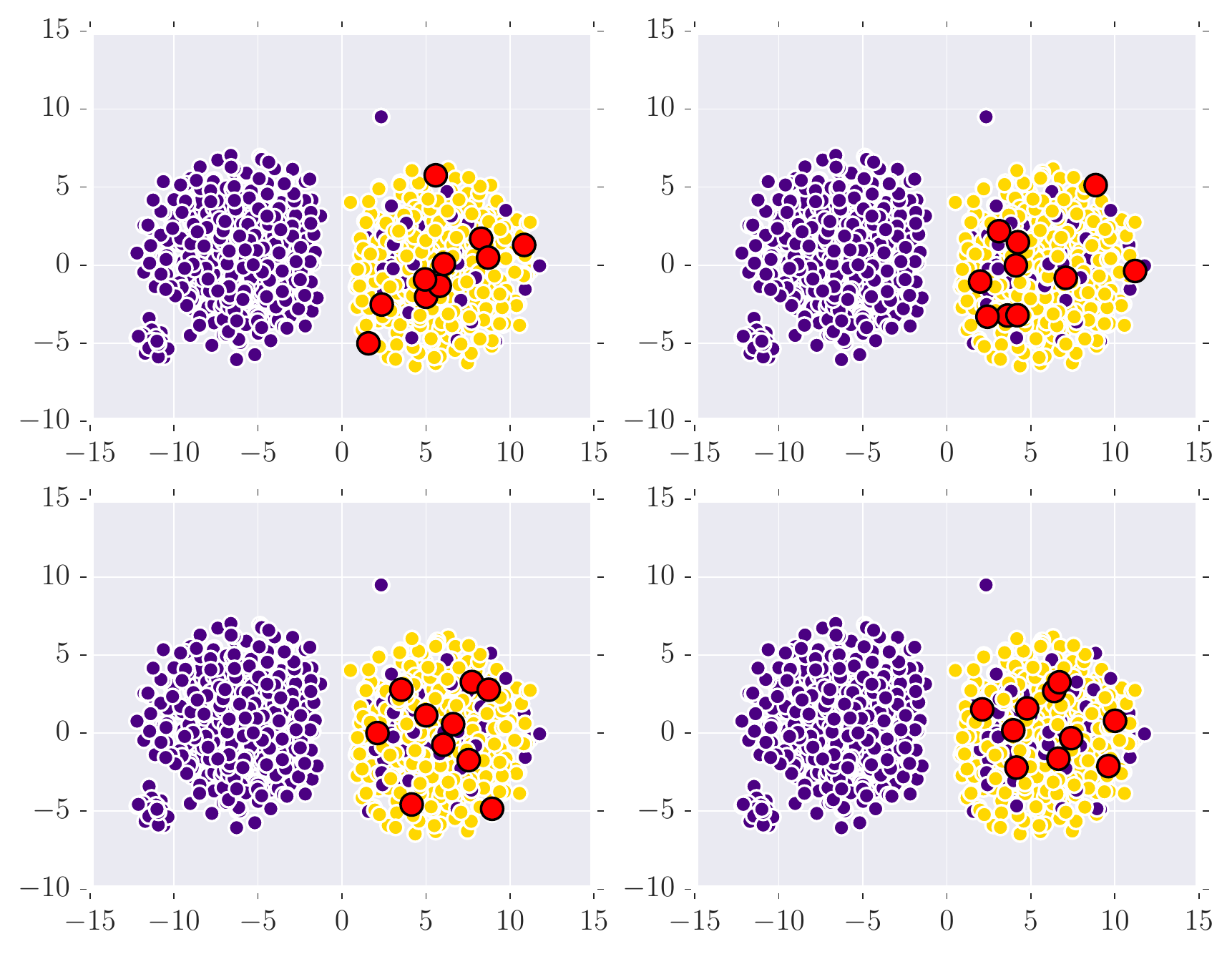}}
    \hfil
    \subfloat[DynamicTriad]{ \includegraphics[width=0.32\textwidth]{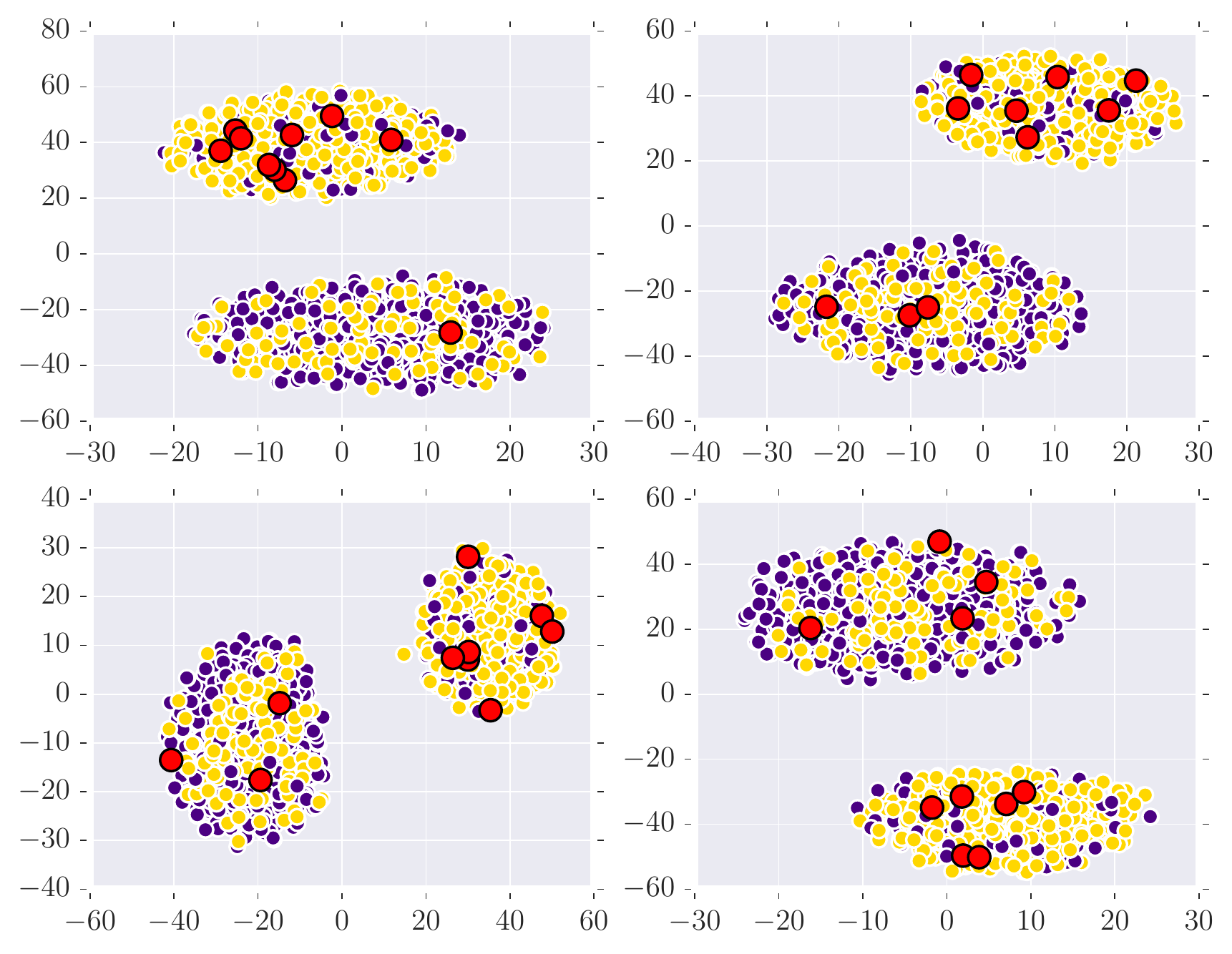}}
       \hfil
    \subfloat[\textit{dyngraph2vecAE}] { \includegraphics[width=0.32\textwidth]{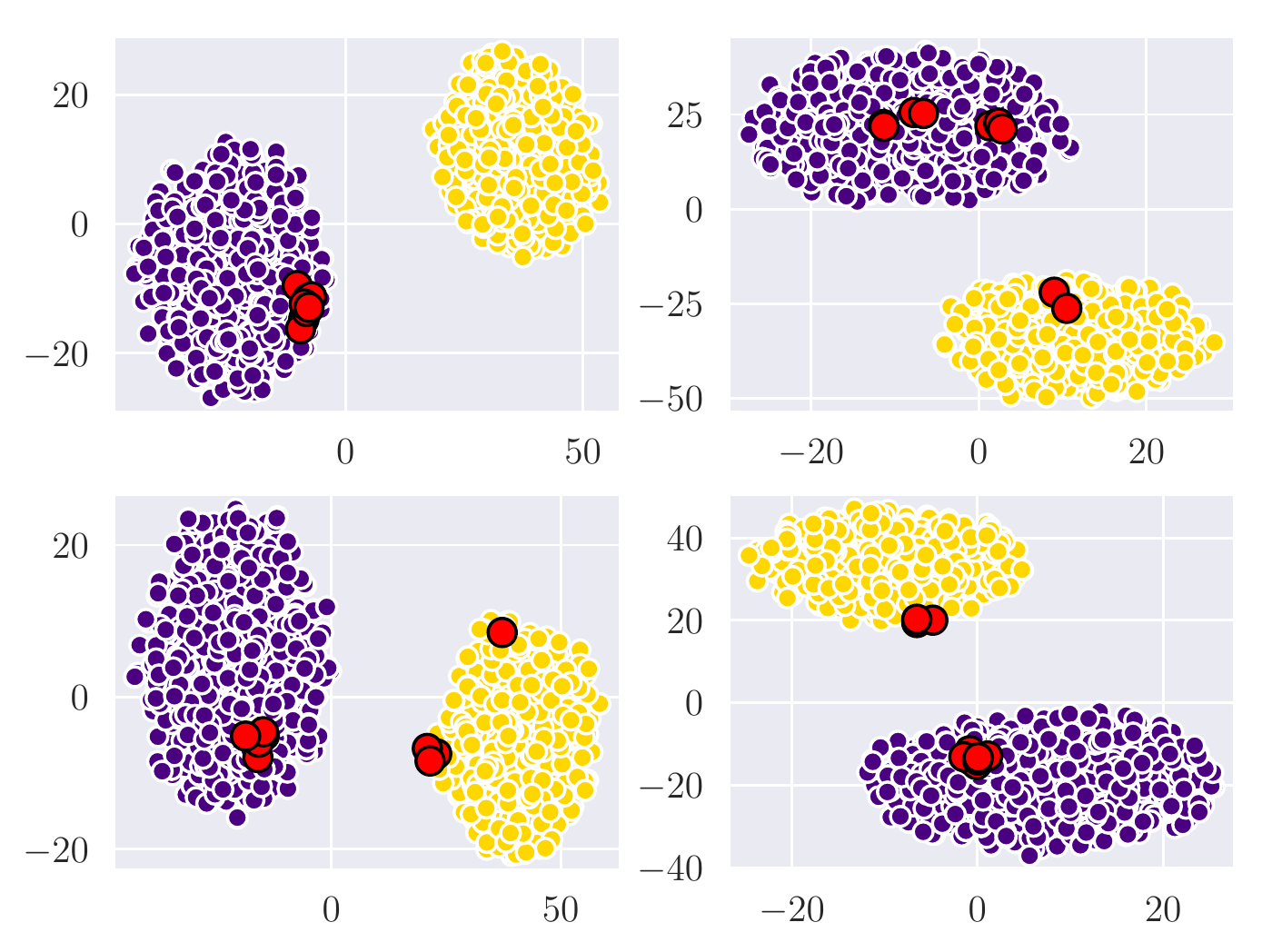}}
     \hfil
    \subfloat[dyngraph2vecRNN]{ \includegraphics[width=0.32\textwidth]{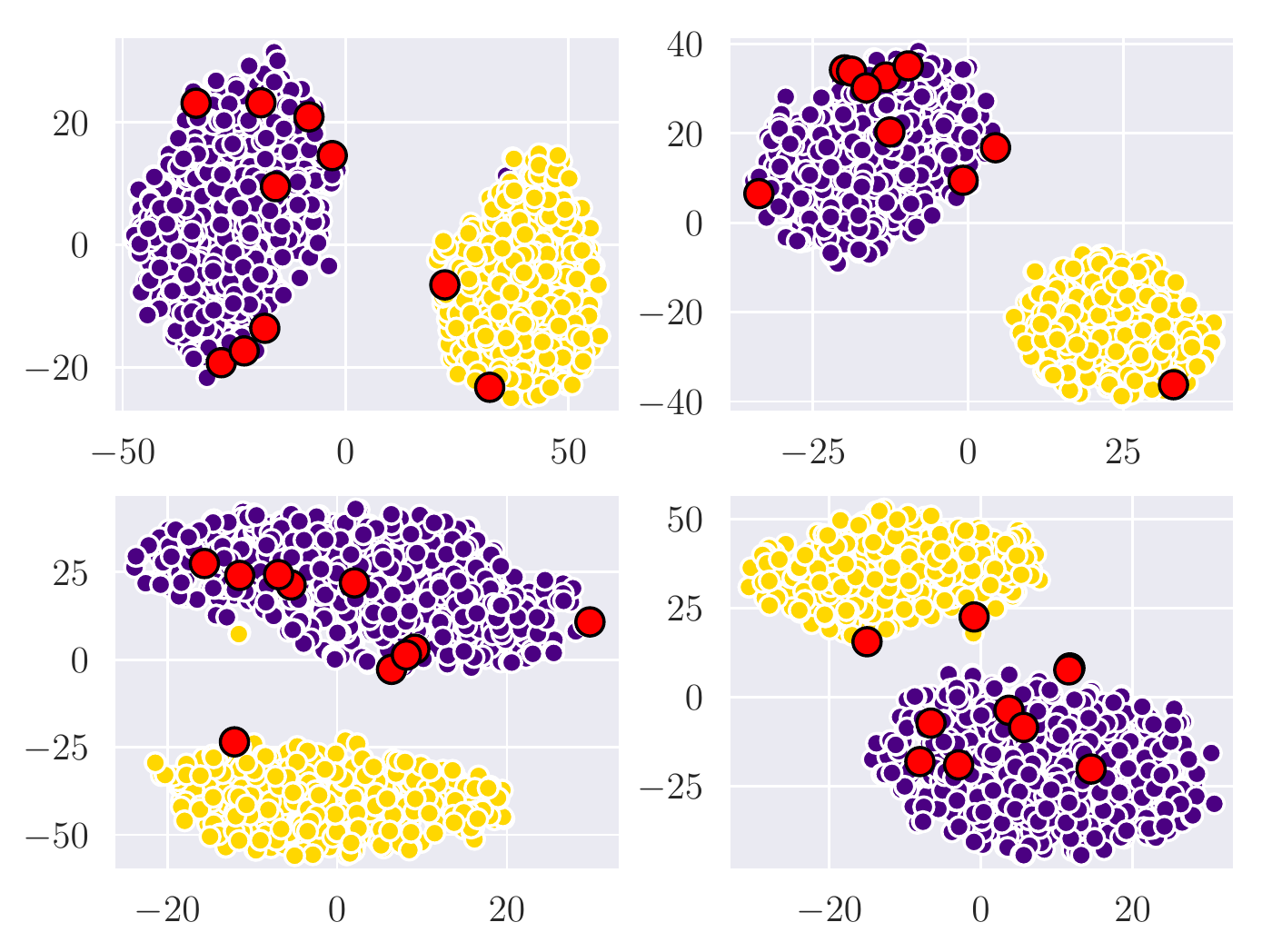}}
     \hfil
    \subfloat[dyngraph2vecAERNN]{ \includegraphics[width=0.32\textwidth]{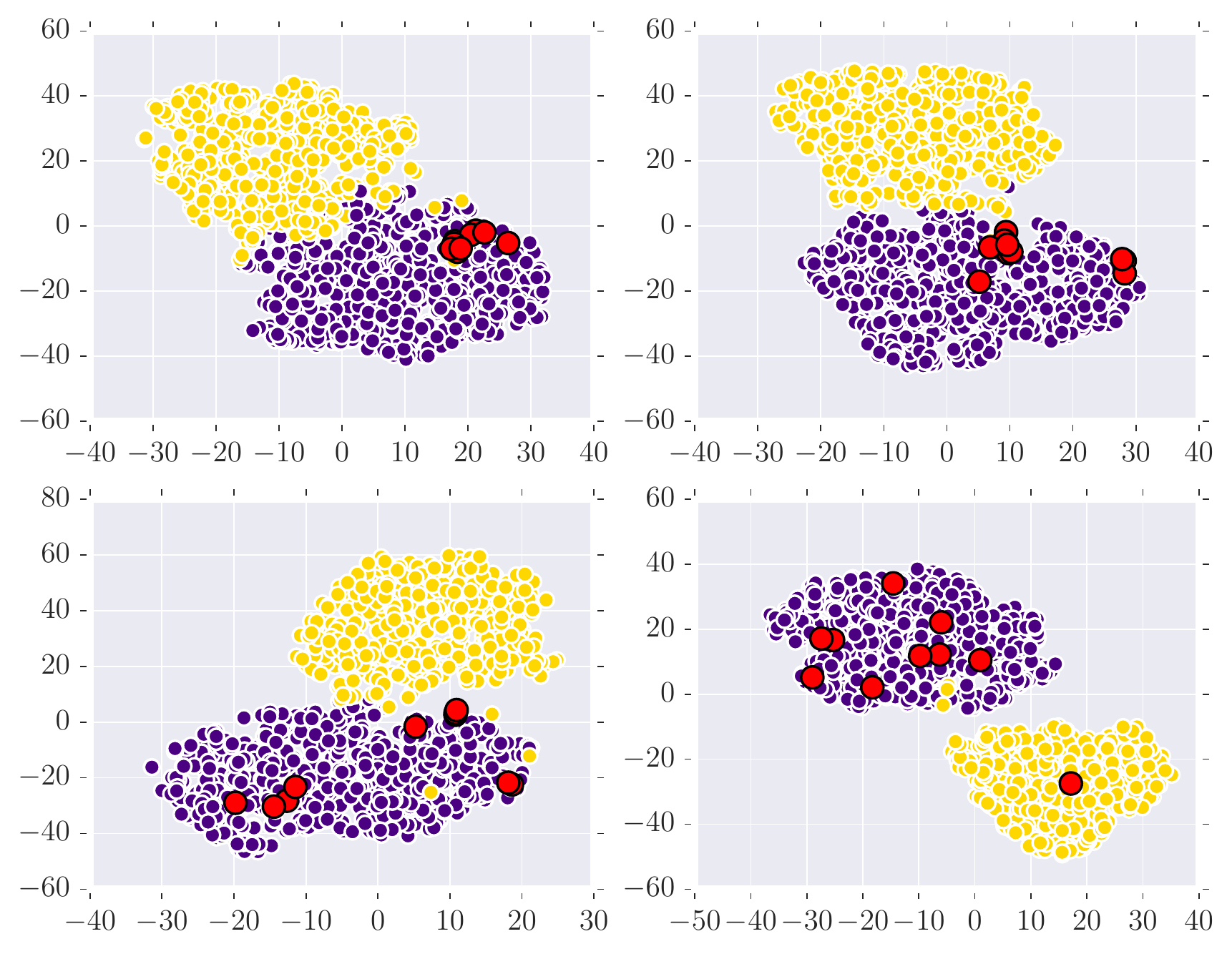}}
  \hfil
    \caption{Motivating example of network evolution - community shift.}
    \label{fig:motivation_results}
\end{adjustbox}
\end{figure*}

\subsection{Dynamic Graph Embedding}
Embedding dynamic graphs is an emerging topic still under investigation.
Some methods have been proposed to extend static graph embedding approaches by adding regularization~\cite{zhu2016scalable,zhang2017timers}.
DynGEM~\cite{goyal2017dyngem} uses the learned embedding from previous time step graphs to initialize the current time step embedding.
Although it does not explicitly use regularization, such initialization implicitly keeps the new embedding close to the previous.
DynamicTriad~\cite{zhou2018dynamic} relaxes the temporal smoothness assumption but only considers patterns spanning two-time steps.
\textcolor{black}{TIMERS~\cite{zhang2017timers} incrementally updates the embedding using incremental Singular Value Decomposition (SVD) and reruns SVD when the error increases above a threshold.
\begin{sloppypar}
DYLINK2VEC~\cite{rahman2018dylink2vec} learns embedding of links (node pairs instead of nodes) and uses temporal functions to learn patterns over time.
\end{sloppypar}
Link embedding renders the method non-scalable for graphs with high density.}
Our model uses recurrent layers to learn temporal patterns over long sequences of graphs and multiple fully connected layer to capture intricate patterns at each time step.

\textcolor{black}{
\subsection{Dynamic Link Prediction}
Several methods have been proposed on dynamic link prediction without emphasis on graph embedding.
Many of these methods use probabilistic non-parametric approaches~\cite{sarkar2012nonparametric,yang2016learning}.
NonParam~\cite{sarkar2012nonparametric} uses kernel functions to model the dynamics of individual node features influenced by the neighbor features.
Another class of algorithms uses matrix and tensor factorizations to model link dynamics~\cite{dunlavy2011temporal,ma2018graph}.
Further, many dynamic link prediction models have been proposed for specific applications including recommendation systems~\cite{talasu2017link} and attributed graphs~\cite{li2018streaming}.
These methods often have assumptions about the inherent structure of the network and require node attributes.
Our model, however, extends the traditional graph embedding framework to capture network dynamics.
}

\section{Motivating Example}\label{sec:motivation}
We consider a toy example to motivate the idea of capturing network dynamics. Consider an evolution of graph $G$, $\mathcal{G} = \lbrace G_1, .., G_T\rbrace$, where $G_t$ represents the state of graph at time $t$. {\color{black} The initial graph $G_1$ is generated using the Stochastic Block Model~\cite{wang1987stochastic} with 2 communities (represented by colors indigo and yellow in Figure \ref{fig:motivation_sbm}), each with 500 nodes (the figure shows a total of 50 nodes for ease of visualization)}.  The in-block and cross-block probabilities are set to 0.1 and 0.01 respectively. The evolution pattern can be defined as a three-step process.
{\color{black} In the first step (shown in Figure \ref{fig:motivation_sbm}(a)), we randomly and uniformly select 10 nodes (colored red in Figure \ref{fig:motivation_sbm} which shows 2 of these nodes) from the yellow community.} In step two (shown in Figure \ref{fig:motivation_sbm}(b)), we randomly add 30 edges between each of the selected nodes in step one and random nodes in the Indigo community. This is similar to having more than cross-block probability but less than in-block probability. In step three (shown in Figure \ref{fig:motivation_sbm}(c)), the community membership of the nodes selected in step 2 is changed from yellow to indigo. Similarly, the edges (colored red in Figure \ref{fig:motivation_sbm}) are either removed or added to reflect the cross-block and in-block connection probabilities. Then, for the next time step (shown in Figure \ref{fig:motivation_sbm}(d)), the same three steps are repeated to evolve the graph.
Informally, this can be interpreted as a two-step movement of users from one community to another by initially increasing friends in the other community and subsequently moving to it. 

\vspace{-.5cm}
\textit{\begin{figure}[!h]
\includegraphics[width=0.48\textwidth]{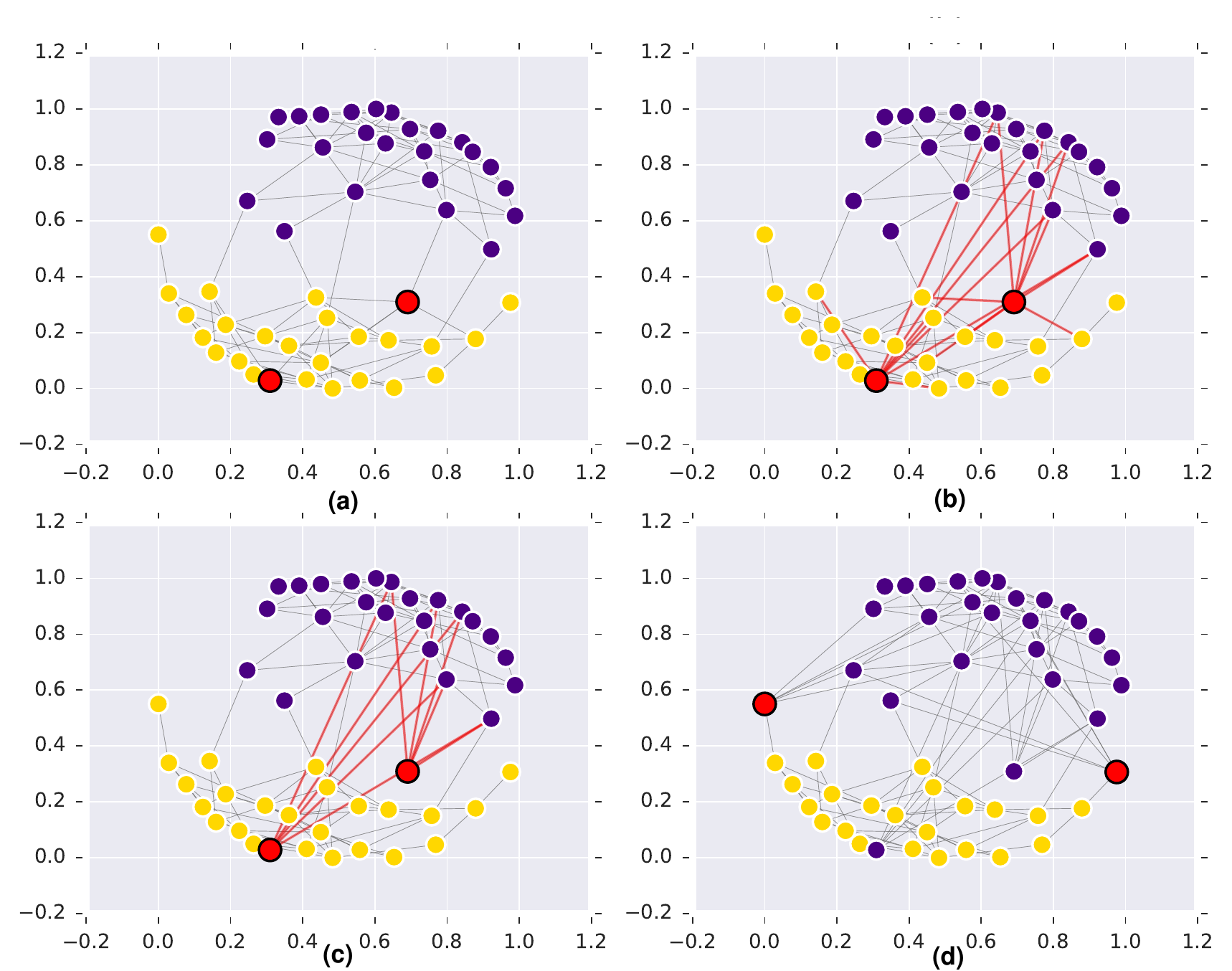}
\vspace{-1em}
\caption{Motivating example of network evolution - community shift (for clarity, only showing 50 of 500 nodes and 2 out 10 migrating nodes). }
\label{fig:motivation_sbm}
\vspace{-0.8em}
\end{figure}}

Our task is to learn the embeddings predictive of the change in community of the 10 nodes. Figure \ref{fig:motivation_results} shows the results of the state-of-the-art dynamic graph embedding techniques (\textit{DynGEM}, \textit{optimalSVD}, and \textit{DynamicTriad}) and the three variations of our model:  \textit{dyngraph2vecAE},  \textit{dyngraph2vecRNN} and  \textit{dyngraph2vecAERNN} (see Methodology Section for the description of the methods). Figure~\ref{fig:motivation_results} shows the embeddings of nodes after the first step of evolution.
The nodes selected for community shift are colored in red. We show the results for 4 runs of the model to ensure robustness. Figure~\ref{fig:motivation_results}(a) shows that DynGEM brings the red nodes closer to the edge of the yellow community but does not move any of the nodes to the other community. Similarly, DynamicTriad results in Figure~\ref{fig:motivation_results}(c) show that it only shifts 1 to 4 nodes to its actual community in the next step. The optimalSVD method in Figure~\ref{fig:motivation_results}(b) is not able to shift any nodes. However, our \textit{dyngraph2vecAE} and  \textit{dyngraph2vecRNN}, and  \textit{dyngraph2vecAERNN} (shown in Figure~\ref{fig:motivation_results}(d-f)) successfully capture the dynamics and move the embedding of most of the 10 selected nodes to the indigo community, keeping the rest of the nodes intact.
This shows that capturing dynamics is critical in understanding the evolution of networks.

\begin{figure*}[!ht]
\centering
\vspace{-0.8em}
\includegraphics[width=0.98\textwidth]{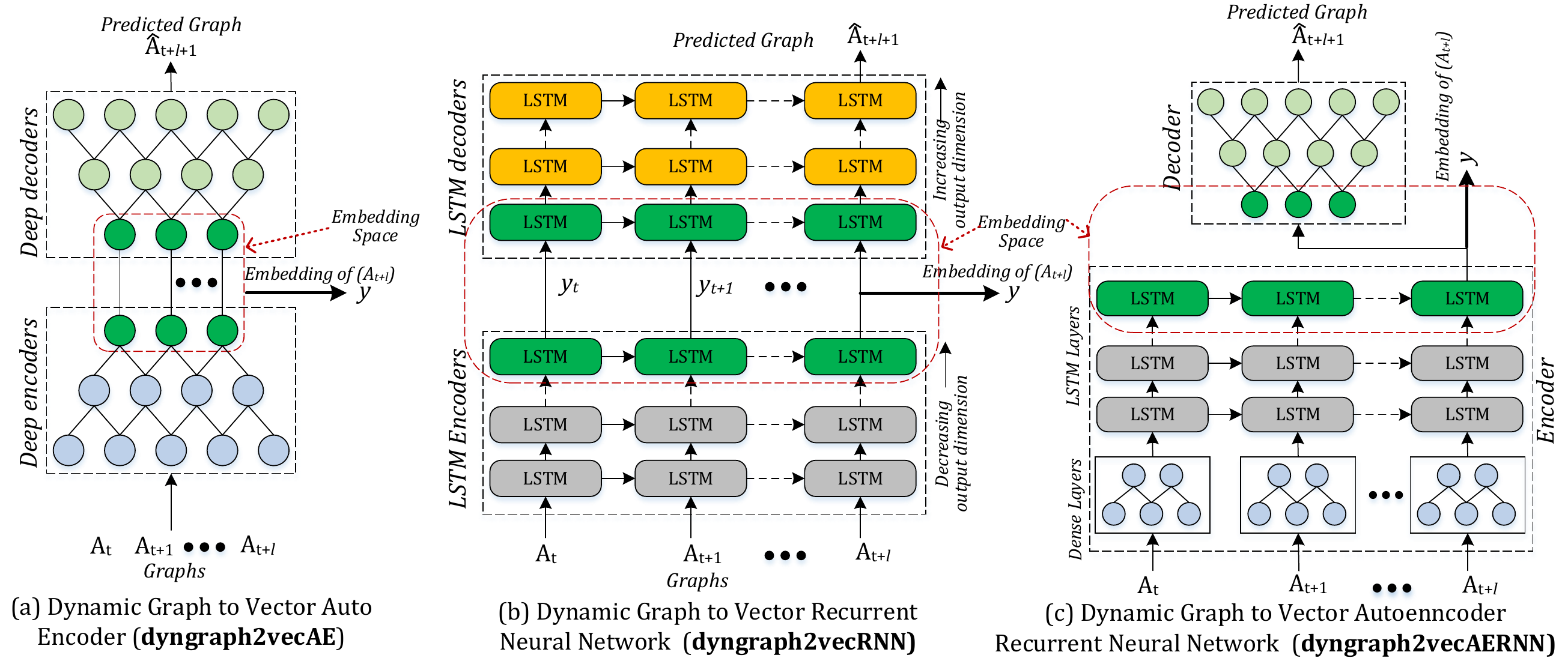}
\vspace{-1em}
\caption{dyngraph2vec architecture variations for dynamic graph embedding.}
\vspace{-0.8em}
\label{fig:architecture}
\end{figure*}


\section{Methodology}\label{sec:methodology}
In this section, we define the problem statement.
We then explain multiple variations of deep learning models capable of capturing temporal patterns in dynamic graphs.
Finally, we design the loss functions and optimization approach.

\subsection{Problem Statement}
Consider a weighted graph $G(V, E)$, with $V$ and $E$ as the set of vertices and edges respectively.
We denote the adjacency matrix of $G$ by $A$, i.e. for an edge $(i, j) \in E$, $A_{ij}$ denotes its weight, else $A_{ij} = 0$.
An evolution of graph $G$ is denoted as $\mathcal{G} = \lbrace G_1, .., G_T\rbrace$, where $G_t$ represents the state of graph at time $t$.

We define our problem as follows: \textit{Given an evolution of graph $G$, $\mathcal{G}$, we aim to represent each node $v$ in a series of low-dimensional vector space $y_{v_1}, \ldots y_{v_t}$, \textcolor{black}{where $y_{v_t}$ is the embedding of node $v$ at time $t$}, by learning mappings $f_t: \{V_1, \ldots, V_t, E_1, \ldots E_t\} \rightarrow \mathbb{R}^d$ and $y_{v_i} = f_i(v_1, \ldots, v_i, E_1, \ldots E_i)$ such that $y_{v_i}$ can capture temporal patterns required to predict $y_{v_{i+1}}$}.
In other words, the embedding function at each time step uses information from graph evolution to capture network dynamics and can thus predict links with higher precision.

\subsection{dyngraph2vec}
\begin{algorithm}[tb]\color{black}
\Fn{dyngraph2vec (Graphs $\mathcal{G} = \lbrace G_1, .., G_T\rbrace$, Dimension $d$, Look back $lb$)}{
 Generate adjacency matrices $\mathcal{A}$ from $\mathcal{G}$\;
 $\vartheta$ $\leftarrow$ RandomInit()\;
 Set $\mathcal{F} = \{(A_{t_u})\}$ for each $u \in V$, for each $t \in \lbrace 1 .. t\rbrace$\;
 \For{$iter=1 \ldots I$}{
     $M$ $\leftarrow$ getArchitectureInput($\mathcal{F}$, $lb$)\;
     Choose $L$ based on the architecture used\;
    $grad$ $\leftarrow$ $\partial L/ \partial \vartheta$\;
    $\vartheta$ $\leftarrow$ UpdateGradient($\vartheta$, $grad$)\;
 }
 $Y$ $\leftarrow$ EncoderForwardPass($G$, $\vartheta$)\;
 return $Y$}
\caption{dyngraph2vec}
\label{alg:elaine}
\end{algorithm}
Our \textit{dyngraph2vec} is a deep learning model that takes as input a set of previous graphs and generates as output the graph at the next time step, thus capturing highly non-linear interactions between vertices at each time step and across multiple time steps. {\color{black}Since the embedding values capture the temporal evolution of the links, it allows us to predict the next time step graph link.}
The model learns the network embedding at time step $t$ by optimizing the following loss function:
\begin{equation}
\begin{split}
 L_{t+l} &= \|(\hat{A}_{t+l+1} - A_{t+l+1}) \odot \mathcal{B}\|_F^2,\\
        &=\|(f(A_{t}, \ldots, A_{t+l}) - A_{t+l+1}) \odot \mathcal{B}\|_F^2. 
\end{split}
\label{eq:loss}
\end{equation}
Here we penalize the incorrect reconstruction of edges at time $t+l+1$ by using the embedding at time step $t+l$.
\textcolor{black}{Minimizing this loss function enforces the parameters to be tuned such that it can capture evolving patterns relations between nodes to predict the edges at a future time step.}
The embedding at time step $t+d$ is a function of the graphs at time steps $t, t+1, \ldots, t+l$ where $l$ is the temporal look back.
We use a weighting matrix $\mathcal{B}$ to weight the reconstruction of observed edges higher than unobserved links as traditionally used in the literature~\cite{Wang2016}.
Here, $\mathcal{B}_{ij} = \beta$ for $(i, j) \in E_{t+l+1}$, else 1, \textcolor{black}{where $\beta$ is a hyperparameter controlling the weight of penalizing observed edges. Note that $\odot$ represents elementwise product.}

\begin{sloppypar}
We propose three variations of our model based on the architecture of deep learning models as shown in Figure~\ref{fig:architecture}: (i)  \textit{dyngraph2vecAE}, (ii)  \textit{dyngraph2vecRNN}, and (iii) \textit{dyngraph2vecAERNN}.
Our three methods differ in the formulation of the function $f(.)$. {\color{black} \textit{dyngraph2vecAE} extends the autoencoders to the dynamic setting in a straightforward manner. Therefore, we overcome the limitations of capturing temporal information and the high number of model parameters through a newly proposed \textit{dyngraph2vecRNN} and \textit{dyngraph2vecAERNN}.}
\end{sloppypar}

\textcolor{black}{A simple way to extend the autoencoders traditionally used to embed static graphs~\cite{Wang2016} to temporal graphs is to add the information about previous $l$ graphs as input to the autoencoder.
This model (\textit{dyngraph2vecAE}) thus uses multiple fully connected layers to model the interconnection of nodes within and across time.}
Concretely, for a node $u$ with neighborhood vector set  $u_{1..t} = [a_{u_t}, \ldots, a_{u_{t+l}}]$, the hidden representation of the first layer is learned as:
\begin{equation}
   y_{u_t}^{(1)} = f_a(W_{AE}^{(1)}u_{1..t} + b^{(1)}),
\end{equation}
where $f_a$ is the activation function, $W_{AE}^{(1)} \in \mathbb{R}^{d^{(1)} \times nl}$ and $d^{(1)}$, $n$ and $l$ are the dimensions of representation learned by the first layer, number of nodes in the graph, and look back,  respectively.
The representation of the $k^{th}$ layer is defined as:
\begin{equation}
   y_{u_t}^{(k)} = f_a(W_{AE}^{(k)}y_{u_t}^{(k-1)} + b^{(k)}).
\end{equation}
Note that \textit{dyngraph2vecAE} has $O(nld^{(1)})$ parameters.
As most real-world graphs are sparse, learning the parameters can be challenging.

\begin{sloppypar}
To reduce the number of model parameters and achieve a more efficient temporal learning, we propose \textit{dyngraph2vecRNN} and \textit{dyngraph2vecAERNN}.
In \textit{dyngraph2vecRNN} we use sparsely connected Long Short Term Memory (LSTM) networks to learn the embedding. LSTM is a type of Recurrent Neural Network (RNN) capable of handling long-term dependency problems. In dynamic graphs, there can be long-term dependencies which may not be captured by fully connected auto-encoders. The hidden state representation of a single LSTM network is defined as:
\end{sloppypar}
\begin{subequations}
\begin{align}
y_{u_t}^{(1)} & = o_{u_t}^{(1)}*\tanh(C_{u_t}^{(1)}) \\
o_{u_t}^{(1)} & = \sigma_{u_t}(W_{RNN}^{(1)}[y_{u_{t-1}}^{(1)},u_{1..t}]+b_o^{(1)}) \\
C_{u_t}^{(1)} & = f_{u_t}^{(1)}*C_{u_{t-1}}^{(1)}+i_{u_t}^{(1)}*\Tilde{C}_{u_t}^{(1)}\\
\Tilde{C}_{u_t}^{(1)} & = \tanh(W_{C}^{(1)}.[y_{u_{t-1}}^{(1)},u_{1..t}+b_{c}^{(1)}])\\
i_{u_t}^{(1)} & = \sigma(W_i^{(1)}.[y_{u_{t-1}}^{(1)},u_{1..t}]+b_{i}^{(1)})\\
f_{u_t}^{(1)} & = \sigma(W_{f}^{(1)}.[y_{u_{t-1}}^{(1)},u_{1..t}+b_{f}^{(1)}])
\end{align}
\end{subequations}
where $C_{u_t}$ represents the cell states of LSTM, $f_{u_t}$ is the value to trigger the forget gate, $o_{u_t}$ is the value to trigger the output gate, $i_{u_t}$ represents the value to trigger the update gate of the LSTM, $\Tilde{C}_{u_t}$ represents the new estimated candidate state, and $b$ represents the biases. There can be $l$ LSTM networks connected in the first layer, where the cell states and hidden representation are passed in a chain from $t-l$ to $t$ LSTM networks. The representation of the $k^{th}$ layer is then given as follows:
\begin{subequations}
\begin{align}
y_{u_t}^{(k)} & = o_{u_t}^{(k)}*\tanh(C_{u_t}^{(k)}) \\
o_{u_t}^{(k)} & = \sigma_{u_t}(W_{RNN}^{(k)}[y_{u_{t-1}}^{(k)},y_{u_t}^{(k-1)}]+b_o^{(k)}) 
\end{align}
\end{subequations}
The problem with passing the sparse neighbourhood vector $u_{1..t} = [a_{u_t}, \ldots, a_{u_{t+l}}]$ of node $u$ to the LSTM network is that the LSTM model parameters (such as the number of memory cells, number of input units, output units, etc.) needed to learn a low dimension representation become large. Rather, the LSTM network may be able to better learn the temporal representation if the sparse neighbourhood vector is reduced to a low dimension representation. To achieve this, we propose a variation of  \textit{dyngraph2vec} model called \textit{dyngraph2vecAERNN}. In  \textit{dyngraph2vecAERNN} instead of passing the sparse neighbourhood vector, we use a fully connected encoder to initially acquire low dimensional hidden representation given as follows:
\begin{equation}
   y_{u_t}^{(p)} = f_a(W_{AERNN}^{(p)}y_{u_t}^{(p-1)} + b^{(p)}).
\end{equation}
where $p$ represents the output  layer of the fully connected encoder. This representation is then passed to the LSTM networks.
\begin{subequations}
\begin{align}
y_{u_t}^{(p+1)} & = o_{u_t}^{(p+1)}*\tanh(C_{u_t}^{(p+1)}) \\
o_{u_t}^{(p+1)} & = \sigma_{u_t}(W_{AERNN}^{(p+1)}[y_{u_{t-1}}^{(p+1)},y_{u_t}^{(p)}]+b_o^{(p+1)}) 
\end{align}
\end{subequations}
Then the hidden representation generated by the LSTM network is passed to a fully connected decoder. 

\subsection{Optimization}
We optimize the loss function defined above to get the optimal model parameters.
By applying the gradient with respect to the decoder weights on equation~\ref{eq:loss}, we get:
\begin{align*}
    \frac{\partial L_{t}}{\partial W_{*}^{(K)}} &= [2 (\hat{A}_{t+1} - A_{t+1}) \odot \mathcal{B}][ \frac{\partial f_a(Y^{(K-1)}W_{*}^{(K)} + b^{(K)})}{\partial W_{*}^{(K)}}],
\end{align*}
where $W_{*}^{(K)}$ is the weight matrix of the penultimate layer for all the three models. For each individual model, we back propagate the gradients based on the neural units to get the derivatives for all previous layers. For the LSTM based \textit{dyngraph2vec} models, back propagation through time is performed to update the weights of the LSTM networks. 

After obtaining the derivatives, we optimize the model using stochastic gradient descent (SGD) \cite{rumelhart1988neurocomputing} with Adaptive Moment Estimation (Adam)\cite{kingma2014adam}.
The algorithm is specified in Algorithm \ref{alg:elaine}.

\section{Experiments}\label{sec:exp}
In this section, we describe the data sets used and establish the baselines for comparison.
Furthermore, we define the evaluation metrics for our experiments and parameter settings.
All the experiments were performed on a 64 bit Ubuntu 16.04.1 LTS system with Intel (R) Core (TM) i9-7900X CPU with 19 processors, 10 CPU cores, 3.30 GHz CPU clock frequency, 64 GB RAM, and two Nvidia Titan X, each with 12 GB memory.

\begin{table}[!htbp]
    \centering
    \renewcommand{\arraystretch}{1.2}
    \caption{Dataset Statistics}
    \label{tab:data_summ}
     \resizebox{0.8\columnwidth}{!}{%
  \begin{tabular}{ l |  c | c | c  }
           \hline \hline
      Name & SBM & Hep-th & AS\\ \hline \hline
      Nodes $n$ &  1000 & 150-14446 & 7716\\ \hline
      Edges $m$ &  56016 &268-48274 & 487-26467\\ \hline
      Time steps $T$ &  10 & 136 & 733\\ \hline
     \hline
  \end{tabular}}
  \vspace{-1.5em}
\end{table}
\subsection{Datasets}
We conduct experiments on two real-world datasets and a synthetic dataset to evaluate our proposed algorithm. {\color{black} We assume that the proposed models are aware of all the nodes, and that no new nodes are introduced in subsequent time steps. Rather, the links between the existing nodes change with a certain temporal pattern.}
The datasets are summarized in Table \ref{tab:data_summ}.\\

\noindent\textbf{Stochastic Block Model (SBM) - community diminishing}: In order to test the performance of various static and dynamic graph embedding algorithms, we generated synthetic SBM data with two communities and a total of 1000 nodes. The cross-block connectivity probability is 0.01 and in-block connectivity probability is set to 0.1. One of the communities is continuously diminished by migrating the 10-20 nodes to the other community. A total of 10 dynamic graphs are generated for the evaluation. {\color{black} Since SBM is a synthetic dataset, there is no notion of time steps.}.\\

\noindent\textbf{Hep-th}~\cite{Gehrke2003}: The first real-world data set used to test the dynamic graph embedding algorithms is the collaboration graph of authors in High Energy Physics Theory conference. The original data set contains abstracts of papers in High Energy Physics Theory conference in the period from January 1993 to April 2003. {\color{black} Hence, the resolution of the time step is one month.} {\color{black} This graph is aggregated over the months.} For our evaluation, we consider the last 50 snapshots of this dataset. From this dataset 2000 nodes are sampled for training and testing the proposed models.\\

\noindent\textbf{Autonomous Systems (AS)}~\cite{leskovec2005graphs}: The second real-world dataset utilized is a communication network of who-talks-to-whom from the BGP (Border Gateway Protocol) logs. The dataset contains 733 instances spanning from November 8, 1997, to January 2, 2000. {\color{black} Hence, the resolution of time step for the AS dataset is one month. However, they are snapshots of each month instead of an aggregation as in Hep-th.} For our evaluation, we consider a subset of this dataset which contains the last 50 snapshots. From this dataset 2000 nodes are sampled for training and testing the proposed models.

\subsection{Baselines}
We compare our model with the following state-of-the-art  static and dynamic graph embedding methods:
\begin{sloppypar}
\begin{itemize}
    \item \textit{Optimal Singular Value Decomposition } (\textbf{OptimalSVD})~\cite{ou2016asymmetric}: It uses the singular value decomposition of the adjacency matrix or its variation (i.e., the transition matrix) to represent the individual nodes in the graph. The low rank SVD decomposition with largest $d$ singular values are then used for graph structure matching, clustering, etc. 
    \item \textit{Incremental Singular Value Decomposition } (\textbf{IncSVD})~\cite{brand2006fast}: It utilizes a perturbation matrix which captures the changing dynamics of the graphs and performs additive modification on the SVD.  
     \item \textit{Rerun Singular Value Decomposition } (\textbf{RerunSVD} or TIMERS)~\cite{zhang2017timers}: It utilizes the incremental SVD to get the dynamic graph embedding, however, it also uses a tolerance threshold to restart the optimal SVD calculation when the incremental graph embedding starts to deviate.  
     \item \textit{Dynamic Embedding using Dynamic Triad Closure Process} (\textbf{dynamicTriad})~\cite{zhou2018dynamic}: It utilizes the triadic closure process to generate a graph embedding that preserves structural and evolution patterns of the graph.
     \item \textit{Deep Embedding Method for Dynamic Graphs} (\textbf{dynGEM})~\cite{goyal2018dyngem}: It utilizes deep auto-encoders to incrementally generate embedding of a dynamic graph at snapshot $t$ by using only the snapshot at time $t-1$. 

\end{itemize}
\end{sloppypar}
    
  

\subsection{Evaluation Metrics}\label{subsec:metrics}

In our experiments, we evaluate our model on link prediction at time step $t+1$ by using all graphs until the time step $t$ . We use Mean Average Precision (MAP) as our metrics. $precision@k$ is the fraction of correct predictions in the top $k$ predictions. It is defined as $P@k= \frac{|E_{pred}(k) \cap E_{gt}|}{k}$, where $E_{pred}$ and $E_{gt}$ are the predicted and ground truth edges respectively. MAP averages the precision over all nodes. It can be written as $\frac{\sum_i AP(i)}{|V|}$ where $AP(i) = \frac{\sum_k precision@k(i) \cdot \mathbb{I}\{E_{pred_i}(k) \in E_{gt_i}\}}{|\{k: E_{pred_i}(k) \in E_{gt_i}\}|}$ and $precision@k(i) = \frac{|E_{pred_i}(1:k) \cap E_{gt_i}|}{k}$.
\textcolor{black}{$P@k$ values are used to test the top predictions made by the model.
MAP values are more robust and average the predictions for all nodes.
High MAP values imply that the model can make good predictions for most nodes.}

\section{Results and Analysis} \label{sec:results}
In this section, we present the performance result of various models for link prediction on different datasets. We train the model on graphs from time step $t$ to $t+l$ where $l$ is the lookback of the model, and predict the links of the graph at time step $t+l+1$. The lookback $l$ is a model hyperparameter. For an evolving graph with $T$ steps, we perform the above prediction from $T/2$ to $T$ and report the average MAP of link prediction. {\color{black} Furthermore, we also present the performance of models when an increasing length of the graph sequence are provided in the training data. Unless explicitly mentioned, for the models consisting of recurrent neural network, a lookback value of 3 is used for the training and testing purpose. }

\begin{figure}[!h]
    \centering
 \includegraphics[width=0.45\textwidth]{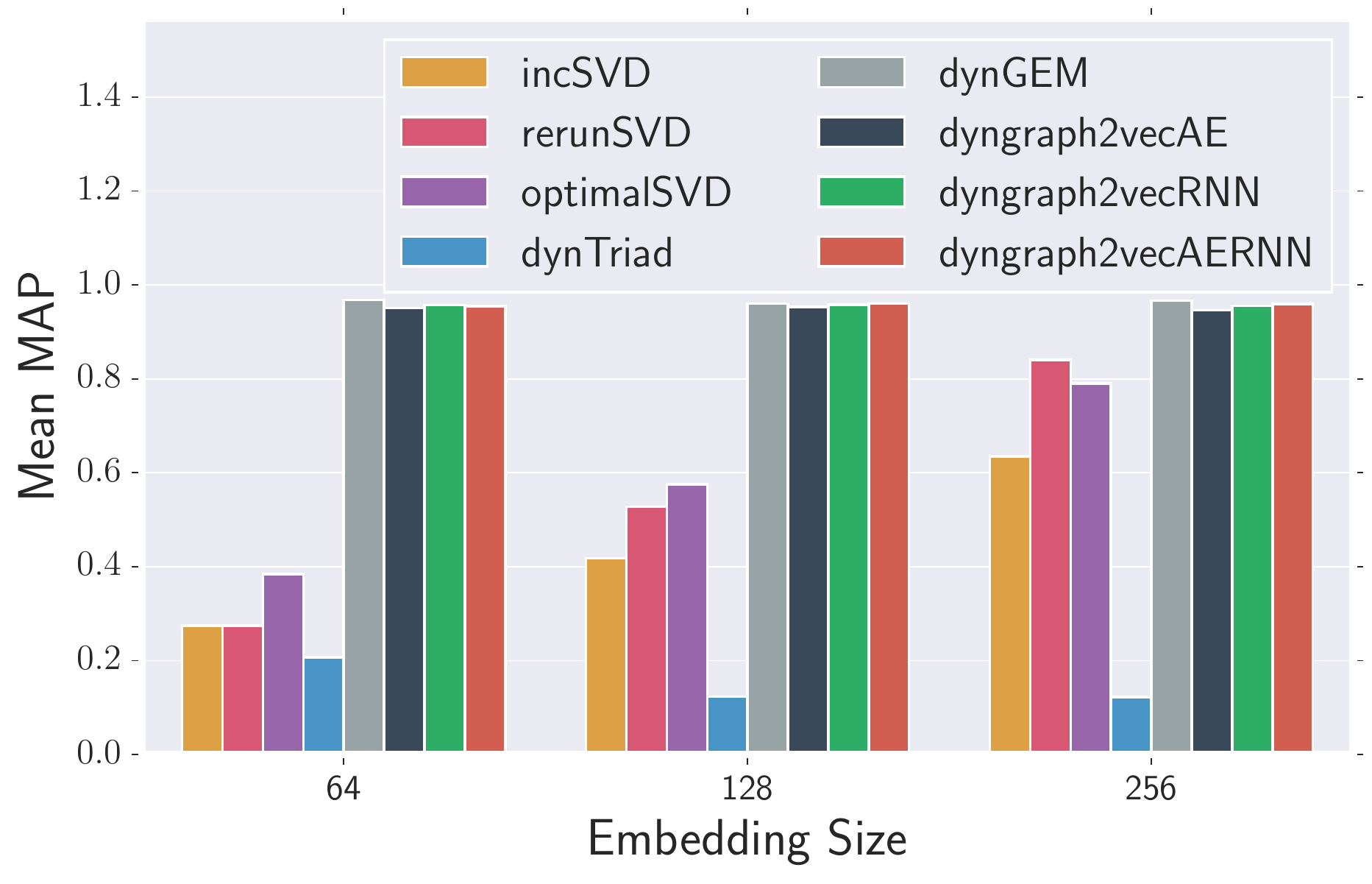}
 \vspace{-1em}
    \caption{MAP values for the SBM dataset.}
    \label{fig:sbm_meanMAP}
     \vspace{-1em}
\end{figure}

\subsection{SBM Dataset}
\begin{sloppypar}
The MAP values for various algorithms with SBM dataset with a diminishing community is shown in Figure \ref{fig:sbm_meanMAP}.
The MAP values shown are for link prediction with embedding sizes \textit{64}, \textit{128} and \textit{256}. This figure shows that our methods \textit{dyngraph2vecAE}, \textit{dyngraph2vecRNN} and \textit{dyngraph2vecAERNN} all have higher MAP values compared to the rest of the base-lines except for \textit{dynGEM}. The \textit{dynGEM} algorithm is able to have higher MAP values than all the algorithms. This is due to the fact that \textit{dynGEM} also generates the embedding of the graph at snapshot $t+1$ using the graph at snapshot $t$.
Since in our SBM dataset the node-migration criteria are introduced only one-time step earlier, the \textit{dynGEM} node embedding technique is able to capture these dynamics. {\color{black} However, the proposed \textit{dyngraph2vec} methods also achieve average MAP values within $\pm$1.5\% of the MAP values achieved by \textit{dynGEM}.}  Notice that the MAP values of SVD based methods increase as the embedding size increases. However, this is not the case for \textit{dynTriad}.
\end{sloppypar}

\begin{figure}[!h]
    \centering
        \vspace{-0.5em}
 \includegraphics[width=0.45\textwidth]{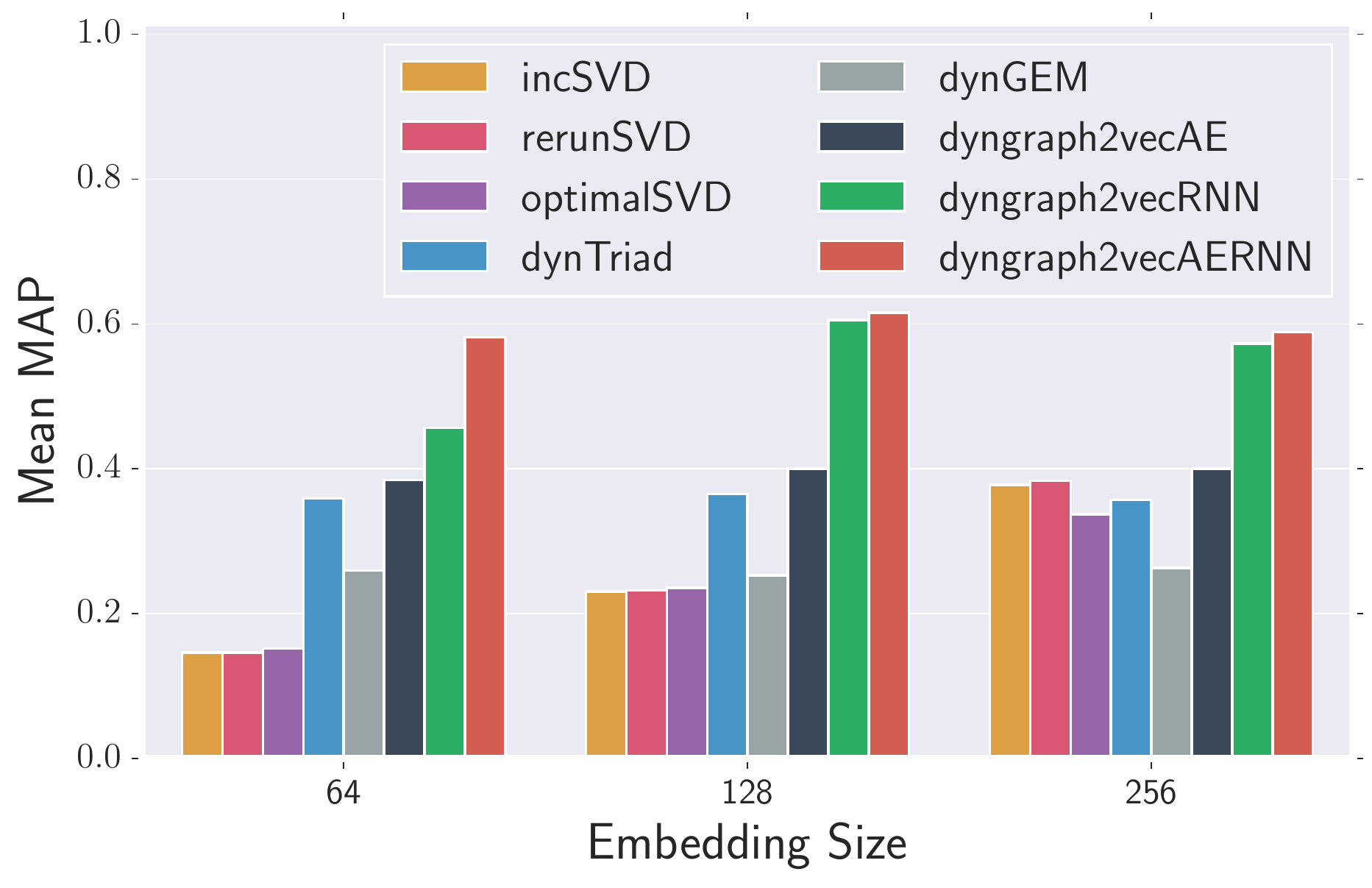}
 \vspace{-1.5em}
    \caption{MAP values for the Hep-th dataset.}
    \label{fig:fig_hepMAP}
    \vspace{-2em}
\end{figure}

\subsection{Hep-th Dataset}
\begin{sloppypar}
The link prediction results for the Hep-th dataset is shown in Figure \ref{fig:fig_hepMAP}. The proposed \textit{dyngraph2vec} algorithms outperform all the other state-of-the-art static and dynamic algorithms. Among the proposed algorithms, \textit{dyngraph2vecAERNN} has the highest MAP values, followed by \textit{dyngraph2vecRNN} and \textit{dyngraph2vecAE}, respectively. The \textit{dynamicTriad} is able to perform better than the SVD based algorithms. Notice that \textit{dynGEM} is not able to have higher MAP values than the \textit{dyngraph2vec} algorithms in the Hep-th dataset.{\color{black}\ Since dyngraph2vec utilizes not only $t-1$ but $t-l-1$ time-steps to predict the link for the time-step $t$, it has higher performance compared to other state-of-the-art algorithms. }
\end{sloppypar}

\begin{figure}[!h]
    \centering
 \includegraphics[width=0.45\textwidth]{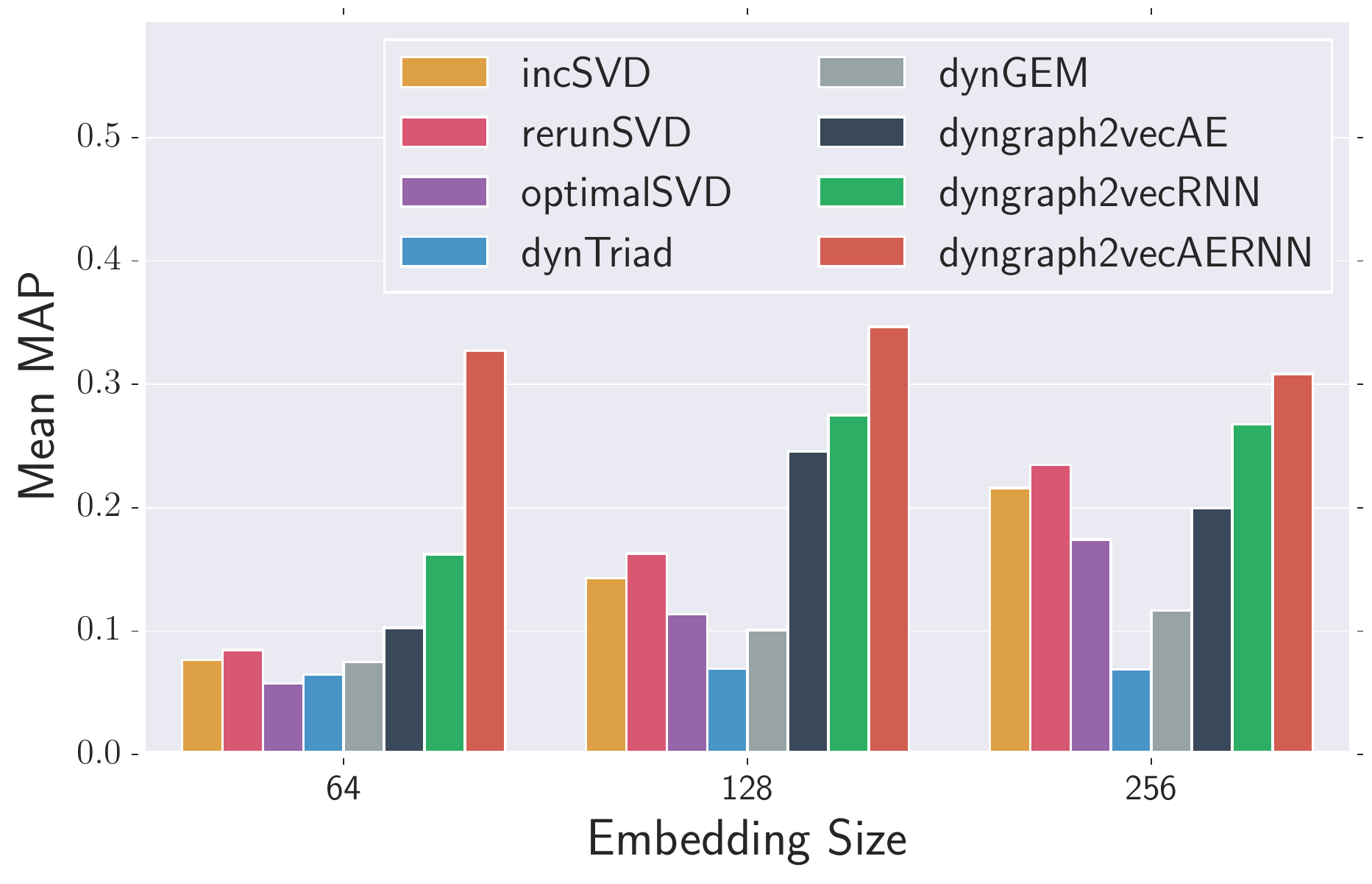}
 \vspace{-1em}
    \caption{MAP values for the AS dataset.}
    \label{fig:fig_AS}
    \vspace{-1em}
\end{figure}

\subsection{AS Dataset}
The MAP value for link prediction with various algorithms for the AS dataset is shown in Figure \ref{fig:fig_AS}.
\textit{dyngraph2vecAERNN} outperforms all the state-of-the-art algorithms.
The algorithm with the second highest MAP score is \textit{dyngraph2vecRNN}.
However, \textit{dyngraph2vecAE} has a higher MAP only with a lower embedding of size 64. SVD methods are able to improve their MAP values by increasing the embedding size. However, they are not able to outperform the \textit{dyngraph2vec} algorithms.

\subsection{MAP exploration}

\begin{sloppypar}
The summary of MAP values for different embedding sizes (64, 128 and 256) for different datasets is presented in Table \ref{tab:summ}. The top three highest MAP values are highlighted in bold. For the synthetic SBM dataset, the top three algorithms with highest MAP values are \textit{dynGEM}, \textit{dyngraph2VecAERNN}, and \textit{dyngraph2vecRNN}, respectively. {\color{black} Since the change pattern for the SBM dataset is introduced only at timestep $t-1$, \textit{dynGEM} is able to better predict the links. The model architecture of \textit{dynGEM} and \textit{dyngraph2vecAE} are only different on what data are fed to train the model. In \textit{dyngraph2vecAE}, we essentially feed more data depending on the size of the lookback. The lookback size increases the model complexity. Since the SBM dataset doesn't have temporal patterns evolving for more than one-time steps, the dyngraph2vec models are only able to achieve comparable but not better result compared to \textit{dynGEM}.   }

\begin{table}[!htbp]
    \centering
    \renewcommand{\arraystretch}{1.2}
    \vspace{-1.5em}
    \caption{Average MAP values over different embedding sizes.}
    
     \vspace{0.5em}
      \resizebox{0.95\columnwidth}{!}{%
  \begin{tabular}{ l ||c|c| l  }
  
           \hline \hline
     &  \multicolumn{3}{c}{Average MAP}\\ \hline 
      Method & SBM & Hep-th & AS\\ \hline \hline
      IncrementalSVD & 0.4421 &  0.2518& 0.1452\\ \hline
      rerunSVD & 0.5474 & 0.2541& 0.1607\\ \hline
      optimalSVD & 0.5831 & 0.2419&0.1152 \\ \hline
      dynamicTriad & 0.1509  & 0.3606&0.0677\\ \hline
      dynGEM & \textbf{0.9648} & 0.2587&0.0975\\ \hline
      \textbf{dyngraph2vec}AE (lb=3) &  0.9500&0.3951 & 0.1825 \\ \hline
      {\color{black}\textbf{dyngraph2vec}AE (lb=5)} &  -&{\color{black}\textbf{0.512}} & {\color{black}\textbf{0.2800}} \\ \hline
      \textbf{dyngraph2vec}RNN (lb=3) & \textbf{0.9567} & 0.5451&0.2350 \\ \hline
      {\color{black}\textbf{dyngraph2vec}RNN } & - & {\color{black}\textbf{0.7290} (lb=8)}&{\color{black}\textbf{0.313} (lb=10)} \\ \hline
      \textbf{dyngraph2vec}AERNN (lb=3) & \textbf{0.9581} & 0.5952& 0.3274\\ \hline  
      {\color{black}\textbf{dyngraph2vec}AERNN } & - & {\color{black}\textbf{0.739 (lb=8)}}& {\color{black}\textbf{0.3801 (lb=10)}}\\ \hline \hline
      \multicolumn{1}{c}{lb = Lookback value}
  \end{tabular}}
  \label{tab:summ}
   \vspace{-1em}
\end{table} 

For the Hep-th dataset, the top three algorithm with highest MAP values are \textit{dyngraph2VecAERNN}, \textit{dyngraph2VecRNN}, and \textit{dyngraph2VecAE}, respectively. {\color{black} In fact, compared to the state-of-the-art algorithm \textit{dynamicTriad}, the proposed models \textit{dyngraph2VecAERNN}(with lookback=8), \textit{dyngraph2VecRNN} (with lookback=8)), and \textit{dyngraph2VecAE}(with lookback=5) obtain $\approx$105\%,  $\approx$102\%, and  $\approx$42\% higher average MAP values, respectively.}

For the AS dataset, the top three algorithm with highest MAP values are \textit{dyngraph2VecAERNN}, \textit{dyngraph2VecRNN}, and \textit{dyngraph2VecAE}, respectively. {\color{black} Compared to the state-of-the-art \textit{rerunSVD} algorithm, the proposed models \textit{dyngraph2VecAERNN}(with lookback=10), \textit{dyngraph2VecRNN} (with lookback=10), and \textit{dyngraph2VecAE} (with lookback=5) obtain $\approx$137\%,  $\approx$95\%, and  $\approx$74\% higher average MAP values, respectively.}

These results show that the dyngraph2vec variants are able to capture the graph dynamics much better than most of the state-of-the-art algorithms in general.
\end{sloppypar}

 \subsection{Hyper-parameter Sensitivity: Lookback }
One of the important parameters for time-series analysis is how much in the past the method looks to predict the future. To analyze the effect of look back on the MAP score we have trained the \textit{dyngraph2Vec} algorithms with various look back values. The embedding dimension is fixed to 128. The look back size is varied from 1 to 10. We then tested the change in MAP values with the real word datasets AS and Hep-th.

\begin{figure}[!h]
    \centering
    \vspace{-1em}
    \includegraphics[width=0.49\textwidth]{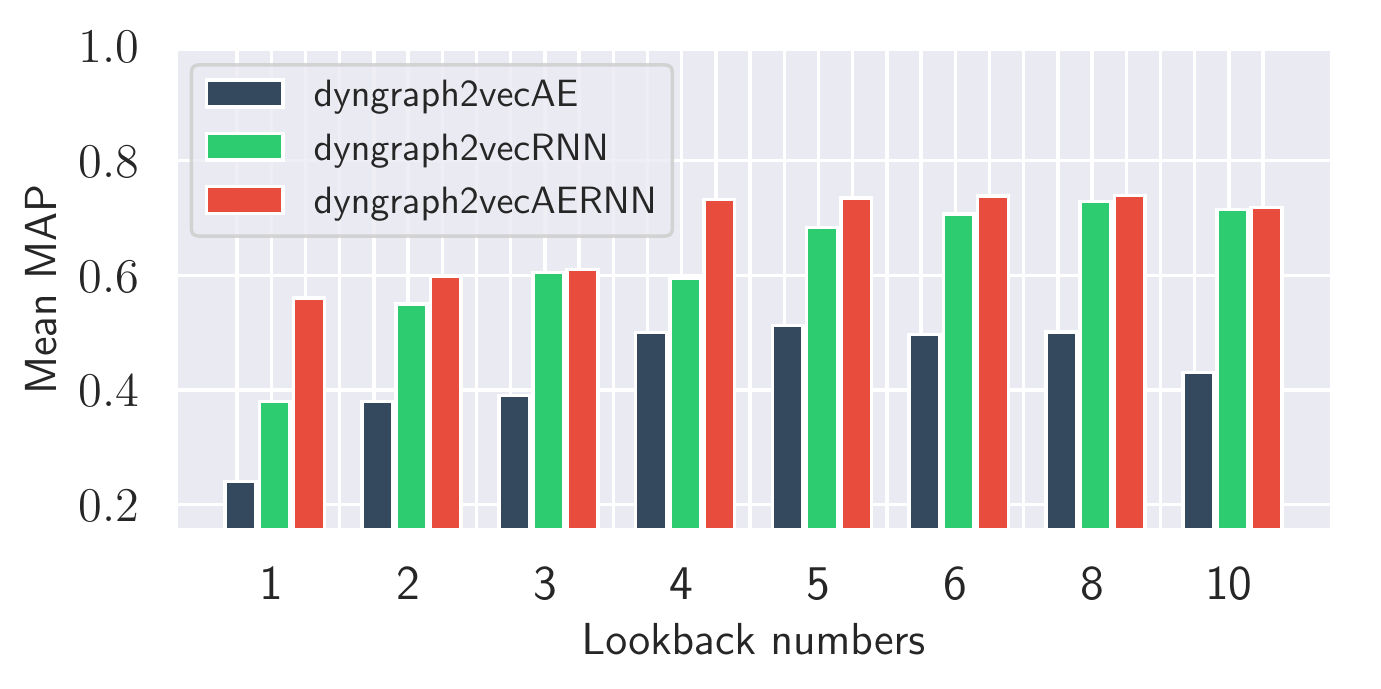}
    \vspace{-2.5em}
    \caption{Mean MAP values for various lookback numbers for Hep-th dataset.}
    \label{fig:lookback_hep}
    \vspace{-0.5em}
\end{figure}

{\color{black} Performance of \textit{dyngraph2Vec} algorithms with various lookback values for the Hep-th dataset is presented in Figure \ref{fig:lookback_hep}. It can be noticed that increasing lookback values consistently increase the average MAP values. Moreover, it is interesting to notice that \textit{dyngraph2VecAE} although has increased in performance until lookback size of 8, its performance is decreased for lookback value of 10. Since it does not have memory units to store the temporal patterns like the recurrent variations, it relies solely on the fully connected dense layers to encode to the pattern. This seems rather ineffective compared to the \textit{dyngraph2VecRNN} and  \textit{dyngraph2vecAERNN} for the Hep-th dataset. The highest MAP values achieved if by \textit{dyngraph2vecAERNN} is 0.739 for the lookback size of 8. }

\begin{figure}[!h]
    \centering
    \vspace{-1em}
    \includegraphics[width=0.49\textwidth]{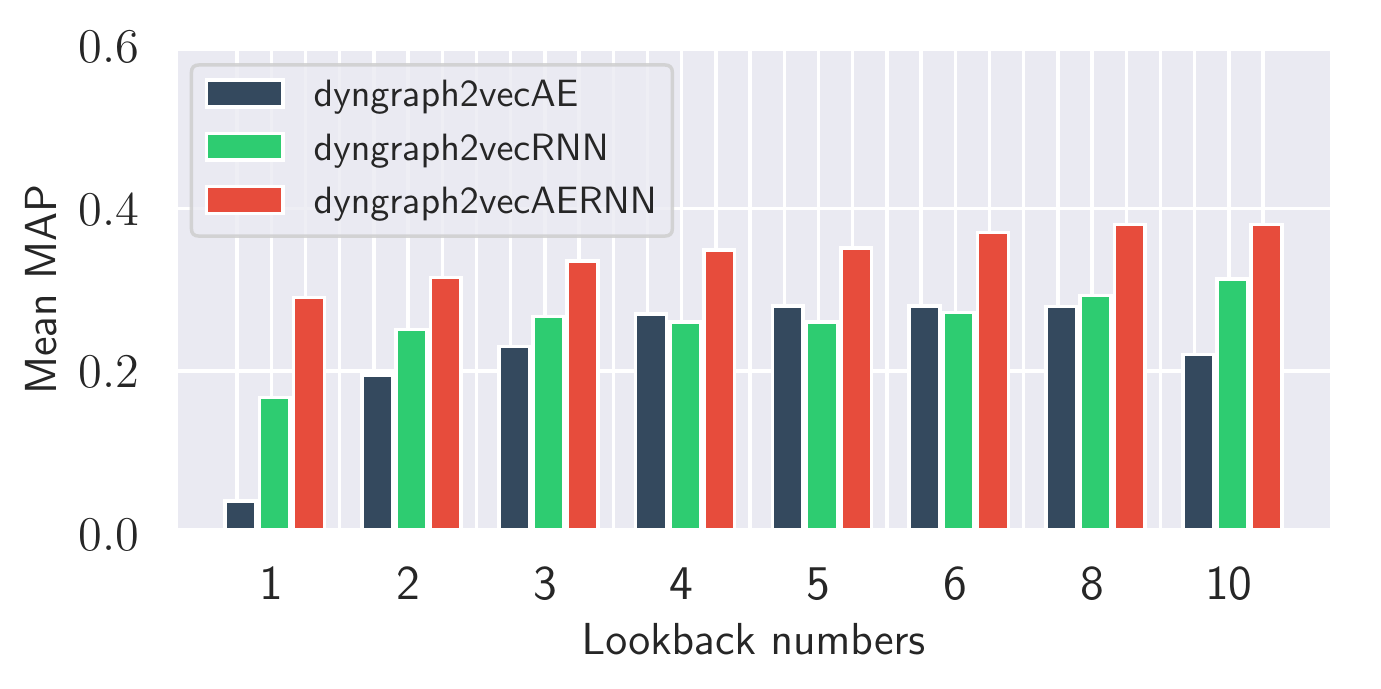}
    \vspace{-2.5em}
    \caption{Mean MAP values for various lookback numbers for AS dataset.}
    \label{fig:lookback_AS}
    \vspace{-1em}
\end{figure}

{\color{black} Similarly, the performance of the proposed models while changing the lookback size for AS dataset is presented in Figure \ref{fig:lookback_AS}. The average MAP values also increase with the increasing lookback size in the AS dataset. The highest MAP value of 0.3801 is again achieved by \textit{dyngraph2vecAERNN} with the lookback size of 10. The \textit{dyngraph2vecAE} model, initially, has comparable and sometimes even higher MAP value with respect to \textit{dyngraph2vecRNN}. However, it can be noticed that for the lookback size of 10, the \textit{dyngraph2vecRNN} outperforms  \textit{dyngraph2vecAE} model consisting of just the fully connected neural networks. In fact, the MAP value does not increase after the lookback size of 5 for  \textit{dyngraph2vecAE}.   }

 {\color{black}
\subsection{Length of training sequence versus MAP value }
In this section, we present the impact of length of graph sequence supplied to the models during training on its performance. In order to conduct this experiment, the graph sequence provided as training data is increased one step at a time. Hence, we use graph sequence of length 1 to $t\in [T, T+1, T+2, T+3, \ldots, T+n]$ to predict the links for graph at time step $t\in [T+1, T+2, \ldots, T+n+1]$, where $T\geq lookback$. The experiment is performed on Hep-th and AS dataset with a fixed lookback size of 8. The total sequence of data is 50 and it is split between 25 for training and 25 for testing. Hence, in the experiment the training data sequence increases from total of 25 sequence to 49 graph sequence. The results in Figure \ref{fig:hep_line} and \ref{fig:AS_line} shows the average MAP values for predicting the links starting the graph sequence at $26^{th}$ to all the way to $50^{th}$ time-step. Where each time step represents a month. 

\begin{figure}[!h]
    \centering
    \vspace{-1em}
    \includegraphics[width=0.49\textwidth]{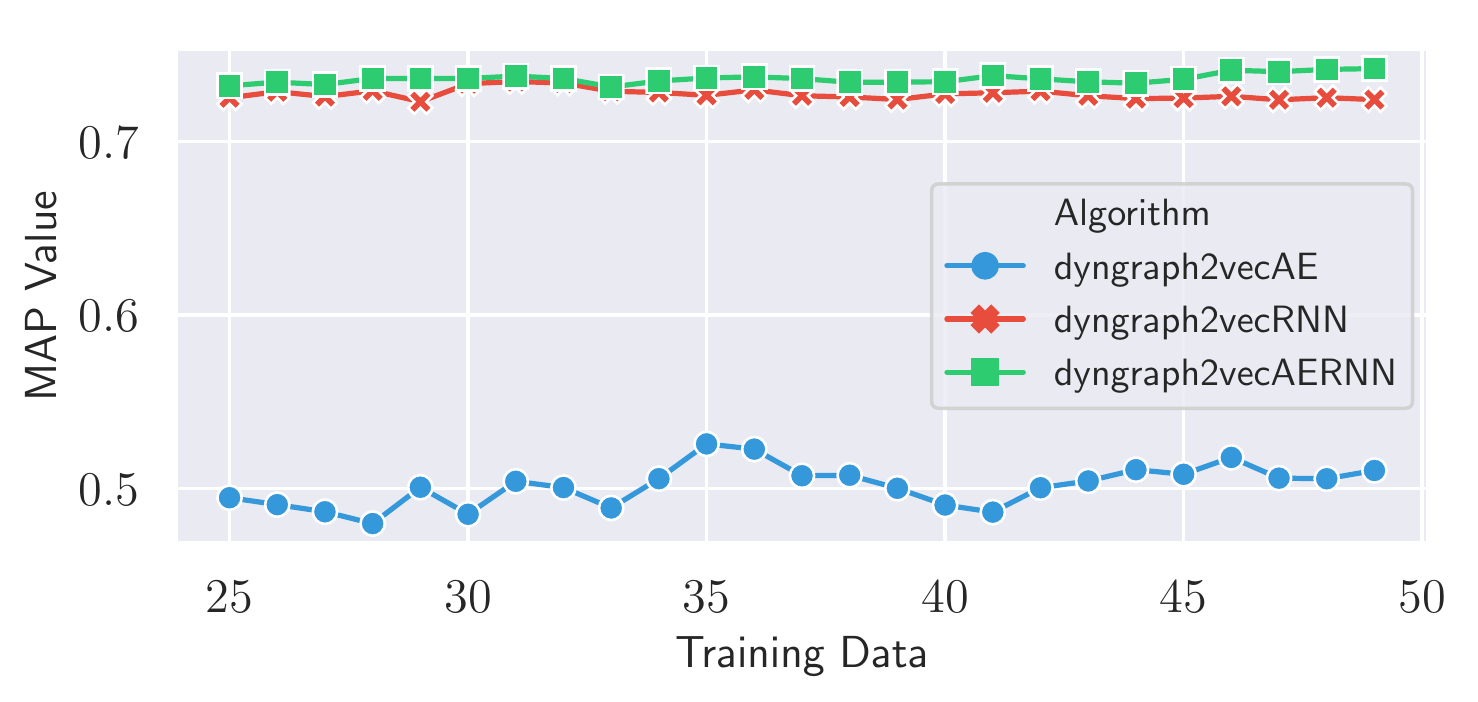}
        \vspace{-2.5em}
    \caption{MAP value with increasing amount of temporal graphs added in the training data for Hep-th dataset (lookback = 8).}
    \label{fig:hep_line}
    \vspace{-0.5em}
\end{figure}

The result of increasing the amount of graph sequence in training data for Hep-th dataset is shown in Figure \ref{fig:hep_line}. It can be noticed that for both the \textit{RNN} and \textit{AERNN} the increasing amount of graph sequence in the data does not drastically increase the MAP value. For, \textit{dyngraph2vecAE} there is a slight increase in the MAP value towards the end. 

\begin{figure}[!h]
    \centering
    \includegraphics[width=0.49\textwidth]{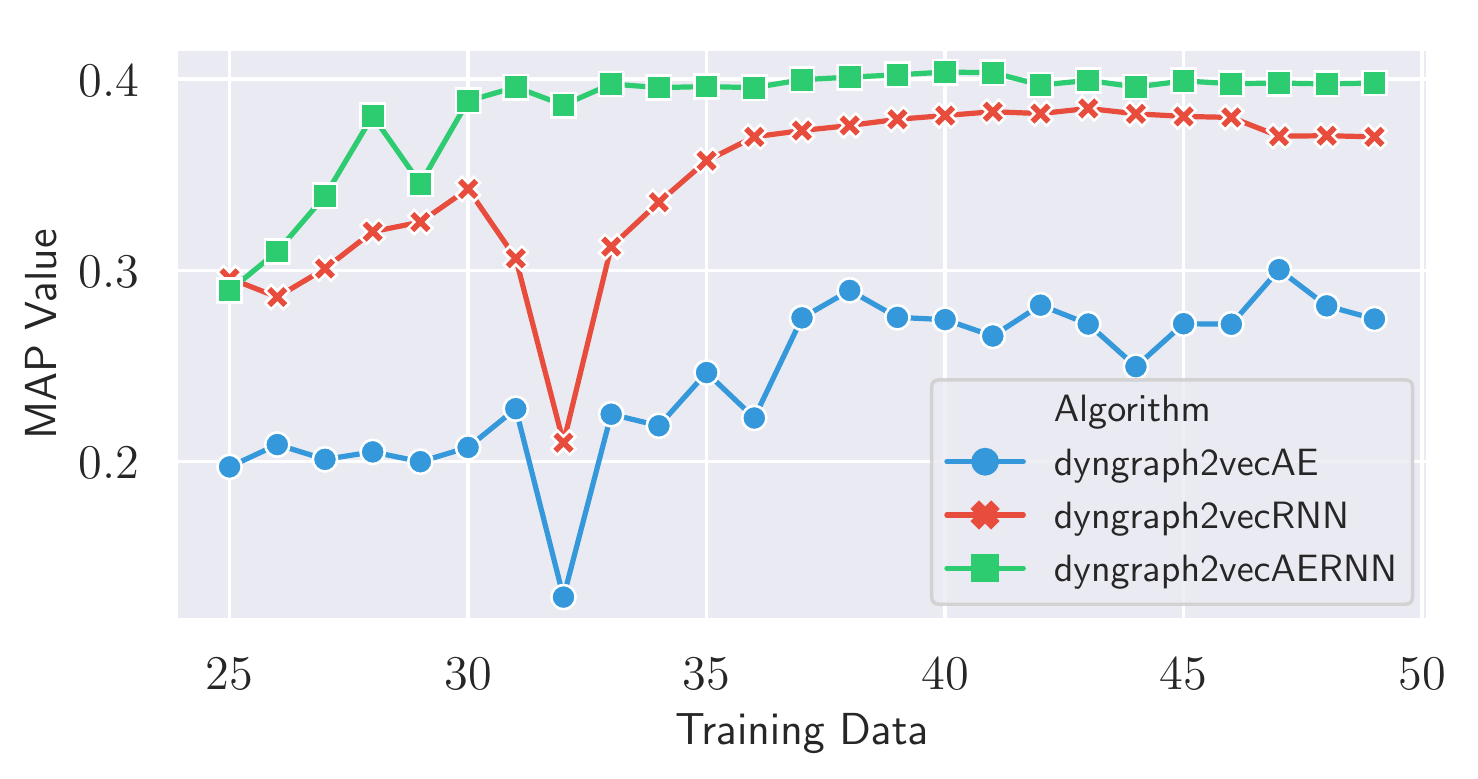}
    \vspace{-2.5em}
    \caption{MAP value with increasing amount of temporal graphs added in the training data for AS dataset (lookback = 8).}
        \label{fig:AS_line}
    \vspace{-1em}
\end{figure}

On the other hand, increasing the amount of graph sequence for the AS dataset during training gives a gradual improvement in link prediction performance in the testing phase. However, they start converging eventually after seeing 80\% (total of 40 graph sequence) of the sequence data.

\section{Discussion}\label{sec:discussion}
\textbf{Model Variation:} It can be observed that among different proposed models, the recurrent variation was capable of achieving higher average MAP values. These architectures are efficient in learning short and long term temporal patterns and provide an edge in learning the temporal evolution of the graphs compared to the fully connected neural networks without recurrent units. 

\textbf{Dataset:} We observe that depending on the dataset, the same model architecture provides different performance. Due to the nature of data, it may have different temporal patterns, periodic, semi-periodic, stationary, etc. Hence, to capture all these patterns, we found out that the models have to be tuned specifically to the dataset. 

\textbf{Sampling:} One of the weakness of the proposed algorithms is that the model size (in terms of the number of weights to be trained) increases based on the size of the nodes considered during the training phase. To overcome this, the nodes have been sampled. Currently, we utilize uniform sampling of the nodes to mitigate this issue. However, we believe that a better sampling scheme that is aware of the graph properties may further improve its performance. 

\textbf{Large Lookbacks:} While it is desirable to test large lookback values for learning the temporal evolution with the current hardware resources, we constantly ran into resource exhausted error with lookbacks greater than 10. (especially for 
}

\section{Future Work}\label{sec:discussion}
\textbf{Other Datasets:} We have validated our algorithms with a synthetic dynamic SBM and two real-world datasets including Hep-th and AS. We leave the test on further datasets as future work. 

\textbf{Hyper-parameters:} Currently, we provided the evaluation of the proposed algorithm with embedding size of 64, 128 and 256. We leave the exhaustive evaluation of the proposed algorithms for broader ranges of embedding size and look back size for future work. 

\textbf{Evaluation:} We have demonstrated the effectiveness of the proposed algorithms for predicting the links of the next time step. However, in dynamic graph networks, there are various evaluations such as node classification that can be performed. We leave them as our future work.

{\color{black} \textbf{Evolving communities:} In real world graphs, communities often grow or shrink in terms of number of nodes per community, and in terms of total number of communities. Using inductive methods to handle such cases is an interesting future work.}

\section{Conclusion}\label{sec:conclusion}
This paper introduced dyngraph2vec, a model for capturing temporal patterns in dynamic networks.
It learns the evolution patterns of individual nodes and provides an embedding capable of predicting future links with higher precision.
We propose three variations of our model based on the architecture with varying capabilities.
The experiments show that our model can capture temporal patterns on synthetic and real datasets and outperform state-of-the-art methods in link prediction. There are several directions for future work: (1) interpretability by extending the model to provide more insight into network dynamics and better understand temporal dynamics; (2) automatic hyperparameter optimization for higher accuracy; and (3) graph convolutions to learn from node attributes and reduce the number of parameters. 

\section*{References}
\balance
\bibliography{mybibfile}

\begin{thebibliography}{10}
\expandafter\ifx\csname url\endcsname\relax
  \def\url#1{\texttt{#1}}\fi
\expandafter\ifx\csname urlprefix\endcsname\relax\def\urlprefix{URL }\fi
\expandafter\ifx\csname href\endcsname\relax
  \def\href#1#2{#2} \def\path#1{#1}\fi

\bibitem{Gehrke2003}
J.~Gehrke, P.~Ginsparg, J.~Kleinberg, Overview of the 2003 kdd cup, ACM SIGKDD
  Explorations 5~(2).

\bibitem{freeman2000visualizing}
L.~C. Freeman, Visualizing social networks, Journal of social structure 1~(1)
  (2000) 4.

\bibitem{theocharidis2009network}
A.~Theocharidis, S.~Van~Dongen, A.~Enright, T.~Freeman, Network visualization
  and analysis of gene expression data using biolayout express3d, Nature
  protocols 4 (2009) 1535--1550.

\bibitem{goyal2018recommending}
P.~Goyal, A.~Sapienza, E.~Ferrara, Recommending teammates with deep neural
  networks, in: Proceedings of the 29th on Hypertext and Social Media, ACM,
  2018, pp. 57--61.

\bibitem{pavlopoulos2008survey}
G.~A. Pavlopoulos, A.-L. Wegener, R.~Schneider, A survey of visualization tools
  for biological network analysis, Biodata mining 1~(1) (2008) 12.

\bibitem{wasserman1994social}
S.~Wasserman, K.~Faust, Social network analysis: Methods and applications,
  Vol.~8, Cambridge university press, 1994.

\bibitem{goyal2017graph}
P.~Goyal, E.~Ferrara,
  \href{http://www.sciencedirect.com/science/article/pii/S0950705118301540}{Graph
  embedding techniques, applications, and performance: A survey},
  Knowledge-Based Systems\href
  {http://dx.doi.org/https://doi.org/10.1016/j.knosys.2018.03.022}
  {\path{doi:https://doi.org/10.1016/j.knosys.2018.03.022}}.
\newline\urlprefix\url{http://www.sciencedirect.com/science/article/pii/S0950705118301540}

\bibitem{Grover2016}
A.~Grover, J.~Leskovec, node2vec: Scalable feature learning for networks, in:
  Proceedings of the 22nd International Conference on Knowledge Discovery and
  Data Mining, ACM, 2016, pp. 855--864.

\bibitem{Ou2016}
M.~Ou, P.~Cui, J.~Pei, Z.~Zhang, W.~Zhu, Asymmetric transitivity preserving
  graph embedding, in: Proc. of ACM SIGKDD, 2016, pp. 1105--1114.

\bibitem{Ahmed2013}
A.~Ahmed, N.~Shervashidze, S.~Narayanamurthy, V.~Josifovski, A.~J. Smola,
  Distributed large-scale natural graph factorization, in: Proceedings of the
  22nd international conference on World Wide Web, ACM, 2013, pp. 37--48.

\bibitem{Perozzi2014}
B.~Perozzi, R.~Al-Rfou, S.~Skiena, Deepwalk: Online learning of social
  representations, in: Proceedings 20th international conference on Knowledge
  discovery and data mining, 2014, pp. 701--710.

\bibitem{Cao2015}
S.~Cao, W.~Lu, Q.~Xu, Grarep: Learning graph representations with global
  structural information, in: KDD15, 2015, pp. 891--900.

\bibitem{Tang2015}
J.~Tang, M.~Qu, M.~Wang, M.~Zhang, J.~Yan, Q.~Mei, Line: Large-scale
  information network embedding, in: Proceedings 24th International Conference
  on World Wide Web, 2015, pp. 1067--1077.

\bibitem{goyal2018embedding}
P.~Goyal, H.~Hosseinmardi, E.~Ferrara, A.~Galstyan, Embedding networks with
  edge attributes, in: Proceedings of the 29th on Hypertext and Social Media,
  ACM, 2018, pp. 38--42.

\bibitem{zhou2018dynamic}
L.~{Zhou}, Y.~{Yang}, X.~{Ren}, F.~{Wu}, Y.~{Zhuang}, {Dynamic Network
  Embedding by Modelling Triadic Closure Process}, in: AAAI, 2018.

\bibitem{goyal2018dyngem}
P.~Goyal, N.~Kamra, X.~He, Y.~Liu, Dyngem: Deep embedding method for dynamic
  graphs, arXiv preprint arXiv:1805.11273.

\bibitem{zhang2017timers}
Z.~Zhang, P.~Cui, J.~Pei, X.~Wang, W.~Zhu, Timers: Error-bounded svd restart on
  dynamic networks, arXiv preprint arXiv:1711.09541.

\bibitem{belkin2001laplacian}
M.~Belkin, P.~Niyogi, Laplacian eigenmaps and spectral techniques for embedding
  and clustering, in: NIPS, Vol.~14, 2001, pp. 585--591.

\bibitem{Wang2016}
D.~Wang, P.~Cui, W.~Zhu, Structural deep network embedding, in: Proceedings of
  the 22nd International Conference on Knowledge Discovery and Data Mining,
  ACM, 2016, pp. 1225--1234.

\bibitem{cao2016deep}
S.~Cao, W.~Lu, Q.~Xu, Deep neural networks for learning graph representations,
  in: Proceedings of the Thirtieth AAAI Conference on Artificial Intelligence,
  AAAI Press, 2016, pp. 1145--1152.

\bibitem{kipf2016variational}
T.~N. Kipf, M.~Welling, Variational graph auto-encoders, arXiv preprint
  arXiv:1611.07308.

\bibitem{kipf2016semi}
T.~N. Kipf, M.~Welling, Semi-supervised classification with graph convolutional
  networks, arXiv preprint arXiv:1609.02907.

\bibitem{bruna2013spectral}
J.~Bruna, W.~Zaremba, A.~Szlam, Y.~LeCun, Spectral networks and locally
  connected networks on graphs, arXiv preprint arXiv:1312.6203.

\bibitem{henaff2015deep}
M.~Henaff, J.~Bruna, Y.~LeCun, Deep convolutional networks on graph-structured
  data, arXiv preprint arXiv:1506.05163.

\bibitem{zhu2016scalable}
L.~Zhu, D.~Guo, J.~Yin, G.~Ver~Steeg, A.~Galstyan, Scalable temporal latent
  space inference for link prediction in dynamic social networks, IEEE
  Transactions on Knowledge and Data Engineering 28~(10) (2016) 2765--2777.

\bibitem{goyal2017dyngem}
P.~Goyal, N.~Kamra, X.~He, Y.~Liu, Dyngem: Deep embedding method for dynamic
  graphs, in: IJCAI International Workshop on Representation Learning for
  Graphs, 2017.

\bibitem{rahman2018dylink2vec}
M.~Rahman, T.~K. Saha, M.~A. Hasan, K.~S. Xu, C.~K. Reddy, Dylink2vec:
  Effective feature representation for link prediction in dynamic networks,
  arXiv preprint arXiv:1804.05755.

\bibitem{sarkar2012nonparametric}
P.~Sarkar, D.~Chakrabarti, M.~Jordan, Nonparametric link prediction in dynamic
  networks, arXiv preprint arXiv:1206.6394.

\bibitem{yang2016learning}
S.~Yang, T.~Khot, K.~Kersting, S.~Natarajan, Learning continuous-time bayesian
  networks in relational domains: A non-parametric approach, in: Thirtieth AAAI
  Conference on Artificial Intelligence, 2016.

\bibitem{dunlavy2011temporal}
D.~M. Dunlavy, T.~G. Kolda, E.~Acar, Temporal link prediction using matrix and
  tensor factorizations, ACM Transactions on Knowledge Discovery from Data
  (TKDD) 5~(2) (2011) 10.

\bibitem{ma2018graph}
X.~Ma, P.~Sun, Y.~Wang, Graph regularized nonnegative matrix factorization for
  temporal link prediction in dynamic networks, Physica A: Statistical
  mechanics and its applications 496 (2018) 121--136.

\bibitem{talasu2017link}
N.~Talasu, A.~Jonnalagadda, S.~S.~A. Pillai, J.~Rahul, A link prediction based
  approach for recommendation systems, in: 2017 international conference on
  advances in computing, communications and informatics (ICACCI), IEEE, 2017,
  pp. 2059--2062.

\bibitem{li2018streaming}
J.~Li, K.~Cheng, L.~Wu, H.~Liu, Streaming link prediction on dynamic attributed
  networks, in: Proceedings of the Eleventh ACM International Conference on Web
  Search and Data Mining, ACM, 2018, pp. 369--377.

\bibitem{wang1987stochastic}
Y.~J. Wang, G.~Y. Wong, Stochastic blockmodels for directed graphs, Journal of
  the American Statistical Association 82~(397) (1987) 8--19.

\bibitem{rumelhart1988neurocomputing}
D.~E. Rumelhart, G.~E. Hinton, R.~J. Williams, Neurocomputing: Foundations of
  research, JA Anderson and E. Rosenfeld, Eds (1988) 696--699.

\bibitem{kingma2014adam}
D.~Kingma, J.~Ba, Adam: A method for stochastic optimization, arXiv preprint
  arXiv:1412.6980.

\bibitem{leskovec2005graphs}
J.~Leskovec, J.~Kleinberg, C.~Faloutsos, Graphs over time: densification laws,
  shrinking diameters and possible explanations, in: Proceedings of the
  eleventh ACM SIGKDD international conference on Knowledge discovery in data
  mining, ACM, 2005, pp. 177--187.

\bibitem{ou2016asymmetric}
M.~Ou, P.~Cui, J.~Pei, Z.~Zhang, W.~Zhu, Asymmetric transitivity preserving
  graph embedding, in: Proceedings of the 22nd ACM SIGKDD international
  conference on Knowledge discovery and data mining, ACM, 2016, pp. 1105--1114.

\bibitem{brand2006fast}
M.~Brand, Fast low-rank modifications of the thin singular value decomposition,
  Linear algebra and its applications 415~(1) (2006) 20--30.

\end{thebibliography}

\end{document}